\documentclass[]{aa}  

\usepackage{graphicx}
\usepackage{txfonts}
\usepackage[colorlinks=true,
    linkcolor=blue,
    citecolor=blue,
    filecolor=magenta,      
    urlcolor=cyan]{hyperref}
\usepackage{graphicx}	
\usepackage{amsmath}	
\usepackage{amssymb}	
\usepackage{comment}
\usepackage{bigstrut}
\usepackage{multirow}
\usepackage{xcolor}
\newcommand{\labsim}[1]{\fontfamily{qcr}\selectfont \textbf{#1}}

\begin{document}  
   \title{On the connection between AGN radiative feedback and massive black hole spin}
   \titlerunning{AGN feedback and MBH spin}
   \authorrunning{F. Bollati et al.}

   \author{F. Bollati
          \inst{1,2} \fnmsep\thanks{fbollati@uninsubria.it}
          \and
          A. Lupi\inst{1,2,3}
          \and 
          M. Dotti\inst{3,2,4}
          \and 
          F. Haardt\inst{1,2,4}
          }

   \institute{
              DiSAT, Universit\`a degli Studi dell'Insubria, via Valleggio 11, I-22100 Como, Italy       
         \and
              INFN, Sezione di Milano-Bicocca, Piazza della Scienza 3, I-20126 Milano, Italy
         \and    
              Dipartimento di Fisica G. Occhialini, Universit\`a di Milano-Bicocca, Piazza della Scienza 3, I-20126 Milano, Italy
         \and 
              INAF, Osservatorio Astronomico di Brera, Via E. Bianchi 46, I-23807 Merate, Italy
             }
    \date{Received XXX; accepted YYY}

\abstract{
We present a novel implementation for active galactic nucleus (AGN) feedback 
through ultra-fast winds in the code \textsc{gizmo}. Our feedback recipe accounts for the angular dependence of radiative feedback upon black hole spin. 
We self-consistently evolve in time i) the gas accretion process from resolved scales to a smaller scale, unresolved (sub-grid) AGN disc, ii) the evolution of the spin of the massive black hole (MBH), iii) the injection of AGN-driven winds into the resolved scales, and iv) the spin-induced anisotropy of the overall feedback process. We test our implementation by following the
propagation of the wind-driven outflow into an homogeneous
medium, and we compare the results against simple analytical
models. Then, we consider an isolated galaxy setup, thought
to be formed from the collapse of a spinning gaseous halo, and
there we study the impact of the AGN feedback on the evolution of the MBH and
the of the host galaxy. 
We find that: i) AGN feedback limits the gas inflow that powers the MBH, with a consequent weak impact on the host galaxy characterized by a star formation (SF) suppression of about a factor of two in the nuclear ($\lesssim$ kpc) region; ii) the impact of AGN feedback on the host galaxy and on
MBH growth is primarily determined by the AGN luminosity, rather than by its angular pattern set by the MBH spin, i.e., more luminous AGNs more efficiently suppress central SF, clearing a wider central cavities and driving outflows with larger semi-opening angles;
iii) the imprint of the angular pattern of the AGN radiation emission manifest in a more clear way at high (i.e., Eddington) accretion rates. At such high rates the more isotropic angular patterns, proper to higher spin values because of light bending, sweeps away gas
in the nuclear region more easily, hence causing a slower MBH mass and spin growths and a higher quenching of the SF.
We argue that the influence of spin-dependent anisotropy of AGN feedback on MBH and galaxy evolution is likely to be relevant in those scenarios characterized by high and prolonged MBH accretion episodes and by high AGN wind-galaxy coupling. Such conditions are more frequently met in galaxy mergers and/or high redshift galaxies.
}

\maketitle



\section{Introduction}

It is now widely accepted that at the center of each galaxy resides a Massive Black Hole \citep[MBH,][]{Koremendy13}. The mechanisms responsible for the very formation of MBHs in the high-redshift Universe is still debated \citep[see, e.g.][and references therein]{Volonteri21}, but there is a consensus that most of their mass accumulates along the cosmic history through gas accretion \citep{Soltan82}, accompanied by substantial energy release observed in emission in luminous Active Galactic Nuclei (AGNs) \citep{LyndenBell69}. Under specific conditions, the energy radiated by AGNs can even surpass the binding energy of the host galaxy \citep{Bower12}. Consequently, AGN radiation has the potential to exert a significant influence on the host galaxy, in the case of an effective interaction with the interstellar medium (ISM), 
a phenomenon known as AGN feedback. This implies that MBHs are not mere spectators in the galaxy formation process; rather, they play a pivotal role in it. 

One of the possible ways in which AGN interact with their host galaxy ISM is through galaxy-wide ($\geq 0.1 - 10$ kpc) energetic outflows and turbulences \citep[e.g.,][]{Nesvadba11,Cicone18, Veilleux20, Fluetsch21}. Theoretically, these outflows may be driven by radiation pressure or AGN-driven winds and
could occur over scales ranging from the accretion disc ($\leq 10^{-2}$ pc) 
to galaxy-wide scales, where they could affect the properties of the
gas in the ISM and beyond. This interplay between AGN feedback and the galaxy ISM results in AGN feedback regulating both the host galaxy star formation and the MBH growth \citep{Harrison17}, suggesting AGN feedback plays a role in the explanation of the tight scaling relations between galaxy properties and the masses of the central MBH \citep{Magorrian98, Ferrarese00, Haring04}.

Due to its relevance, AGN feedback has become an imperative ingredient
in modern theories of galaxy formation to reproduce key observables of galaxy populations 
and it is
routinely incorporated both in semi-analytic \citep{Kauffmann00, Croton06, Henriques15}, and hydrodynamical simulation models \citep{Vogelsberger14, Hirschmann14, Schaye15, Dave19, Weinberger17}. 
Despite this central
importance of MBHs, the physical processes governing
gas accretion and the associated feedback processes are only poorly understood and the modelling in cosmological hydrodynamic simulations (where MBH accretion disc scales cannot be properly resolved) is hence very
sketchy, and typically encapsulated in heuristic sub-grid models.

Sub-grid feedback models in Lagrangian 
codes (based on particles or moving-mesh) typically
fall into two categories: thermal and kinetic energy injection modes. In thermal mode, AGN feedback is modelled by injecting an amount of thermal energy into the neighbour gas particles at a rate that is directly proportional
to the AGN bolometric luminosity \citep{Springel05, DiMatteo05, Costa14}. While some modifications have been introduced, black hole accretion and quasar feedback follow this basic model in most state-of-the-art cosmological simulations \citep{Vogelsberger13, Schaye15, Tremmel16, Weinberger17, Henden18, Lupi19,Lupi22}. On the other hand, in kinetic mode energy is injected in kinetic form into a number of cell/particle neighbours (e.g. \citealt{Choi12,Barai16}, the ``low accretion
mode'' in \citealt{Weinberger17}, \citealt{Alcazar17,Dave19,Sala21}).
Both these methods have some limitations (see  \citealt{Costa20}, section 5.1, for a detailed discussion): i) they fail to reproduce  the correct AGN wind thermalization scale, provided this can be resolved, ii) they are accompanied by a decrease in the resolution
around the accreting black hole, once the neighbour gas particles are driven outwards by the energy injection, and  iii) the injection itself is anisotropic as it follows the mass distribution of such neighbours gas particles.
\cite{Costa20} proposed a novel sub-grid model in which wind mass is explicitly injected along with momentum and energy at a fixed spatial scale across a desired solid angle, independently of the configuration of the gas cells surrounding the black hole. A similar approach is followed by \cite{Torrey20}, which consists in directly spawning wind particles and in ejecting them outward into the MBH surrounding resolved scales. Both these ``wind injection'' approaches do not suffer from the limitations mentioned above. 

An important parameter to be considered when modeling AGN feedback, and more generally MBH evolution, is the MBH spin, owing to the complex non-linear influence spin and feedback have on each other. 
Indeed, on one hand the spin
modulates radiative efficiency of AGN discs, which influences the MBH accretion and the amount of energy released in radiation, and, for thick
radiatively inefficient discs, it regulates the kinetic power and direction of jets \citep{Blandford77}.
 On the other hand, feedback
strongly impacts the MBH spin growth as it affects the gas reservoir that fuels MBH accretion which, together
with MBH mergers, is the main channel for spin evolution \citep{Berti08}. 
In addition to its relevance in the context of AGN feedback, the spin has a strong influence on the gravitational wave emission of merging BHs \citep{Klein16}, and thus also on the expected recoil velocity of the merger remnant \citep{Dotti10}, which make the spin a fundamental parameter to be considered when we aim at understanding the cosmic evolution of MBHs.  Due to its importance, some recent works started to include spin evolution in hydrodynamical simulations \citep{Fiacconi18, Bustamante19, Cenci21, Dubois21}
 and semi-analytical models of galaxy formation \citep{Volonteri05, Fanidakis11,Barausse12, Sesana14}.
These studies, together with observations \citep{Reynolds21}, have shown that the distribution of MBH spins depends on several
quantities, such as host galaxy morphology, MBH mass, and redshift. 

Recently, \cite{Campitiello18} first discussed the influence that the MBH spin has in shaping the angular pattern of the AGN radiation, showing that more rapidly spinning black holes result in more isotropic radiation patterns, as the geodesics of photons emitted in inner region of the disc undergo a stronger gravitational bending. \cite{Ishibashi19} and \cite{Ishibashi20} first discussed the relevance of this effect in the context of radiation pressure-driven outflows in isolated spherical galaxies, by means of semi-analytic models. They showed that AGNs with rapidly spinning MBHs launch quasi-spherical outflows propagating on large scale at all inclination angles, opposite to MBHs with low spin values that produce weaker bipolar outflows driven in the polar direction. As a consequence, \cite{Ishibashi20} argued that AGNs with slowly spinning MBHs should be accompanied by higher obscuration levels and higher accretion rates, being the AGN radiation less prone to remove gas from the disc equatorial plane. Numerical simulations also suggest that the AGN anisotropic radiation can have a dramatic effect on the outflow properties \citep{Williamson19} and MBH pair dynamics \citep{Bollati23}, but in these studies the anisotropy factor remains unconstrained and simply left as a free parameter. 

In this paper, we present a new implementation of AGN radiative feedback in the code \textsc{gizmo} \citep{Hopkins15} that takes into account the spin-dependence of feedback anisotropy.
In this model, accretion from resolved scales onto an unresolved (sub-grid) AGN disc, spin evolution, the injection of AGN winds into resolved scales and their spin-induced anisotropy, are all self-consistently evolved. 
This implementation builds upon existing modules for MBH
accretion and spin evolution \citep{Cenci21} and AGN wind \citep{Torrey20}. Equipped with this model, we investigate the role of AGN wind anisotropy in shaping AGN-driven outflows and the evolution of isolated disc galaxies hosting active MBHs. 
This paper is organized as follows: in section \ref{sec: Theo} we review the spin-dependence of AGN radiation angular pattern and we connect it to the anisotropy of AGN winds. The implementation of this effect in \textsc{gizmo} is presented in section \ref{sec:model}, and section \ref{Sec: tests} shows some tests of this model. We discuss an application of it in the context of isolated disc galaxies in section \ref{sec: isolated galaxy}, and we draw our conclusions in section \ref{sec: Discussion}.

\section{Theoretical Background}\label{sec: Theo}

In this section, we review how the MBH spin influences both the accretion-disc radiative efficiency and the angular patter of the emitted radiation, and we show how this reflects on the properties of AGN radiation-driven winds. Then, we discuss two analytic solutions of outflows driven by such anisotropic winds.    

\subsection{Radiation angular pattern from accretion discs}\label{ang pattern}
The MBH spin is characterized by the dimensionless spin parameter $a = cJ_\bullet / G M_\bullet ^2$, where $J_\bullet$ is the MBH angular momentum magnitude, $M_\bullet$ the MBH mass, $c$ the speed of light in vacuum and $G$ the gravitational constant. 
The MBH spin determines the location of the innermost stable circular orbit (ISCO) \citep{Bardeen72} of accretion discs and hence their radiative efficiency $\eta$, i.e., the fraction of rest-mass energy accreting onto the BH $\dot{M}_\textrm{acc} c^2$ which is converted in luminosity $L$:
\begin{equation}
    \eta(a) \equiv \frac{L}{\dot{M}_\textrm{acc}c^2}.
    \label{Eq:eta}
\end{equation}
Indeed, the gas in the disc, in order to reach a smaller ISCO, as occurs for higher spin values, needs to dissipate more energy through viscous torques, which results in higher temperatures and higher radiation output.

Recently, \cite{Campitiello18} and \cite{Ishibashi20} pointed out that the BH spin not only influences the amount of energy released in radiation during the accretion process, i.e., the disc luminosity, but also the angular pattern of such radiation. This is a consequence of the spin-dependence of the location of the ISCO and of the relativistic gravitational bending of photons being more effective closer to the BH, i.e., in a stronger gravitational field. Indeed, in the Newtonian case, with straight-lines photon geodesics, the luminosity angular distribution follows a simple cosine-like pattern, with the maximum luminosity observed when the disc is face-on and the minimum in the side-on configuration. If we take into account photon geodesics in full GR, due to the gravitational bending more radiation is capable to reach to observer's eye in the side-on configuration, and this occurs in a way that is sensitive to the BH spin. In particular, for larger BH spin values, the ISCO is located nearer to the BH and therefore the photons emitted by the inner annuli of the disc (the ones that dominate the disc luminosity) experience a stronger gravitational bending, funneling more radiation in the side-on direction, yielding a more isotropic radiation angular pattern. 

\begin{figure}
    \centering
    \includegraphics[scale=0.6]{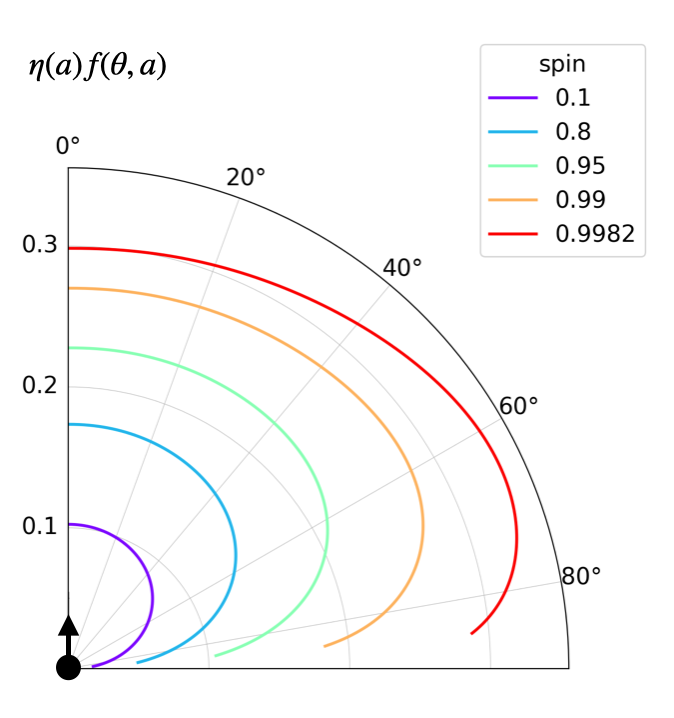}
    \caption{The luminosity angular pattern $\eta(a)f(\theta ;a)$ for different spin values.}
    \label{fig: angular_pattern}
\end{figure}

\cite{Campitiello18} and \cite{Ishibashi20} computed the precise emission pattern numerically by means of the KERRBB model implemented in XSPEC \citep{Li05}. In addition, \cite{Campitiello18} proposed a normalized fitting function $f(\theta;a)$ describing the luminosity angular pattern for different viewing angles $\theta$ and spin parameter $a$, so that 
\begin{equation}
    L(\theta;a) = f(\theta;a)L
    \label{Eq: normalization}
\end{equation}
is the luminosity measured by an observer whose line-of-sight forms an angle $\theta$ with the spin direction.
With the above definition, the optically thick emission from a non-relativistic disc (i.e., when any light bending is neglected) would be described by $f(\theta,a)=f(\theta)=2\cos\theta$. The term $f(\theta,a)\eta(a)$, in the relativistic case, is shown in Fig.~\ref{fig: angular_pattern}.  

\begin{figure}
    \centering
    \includegraphics[scale=0.45]{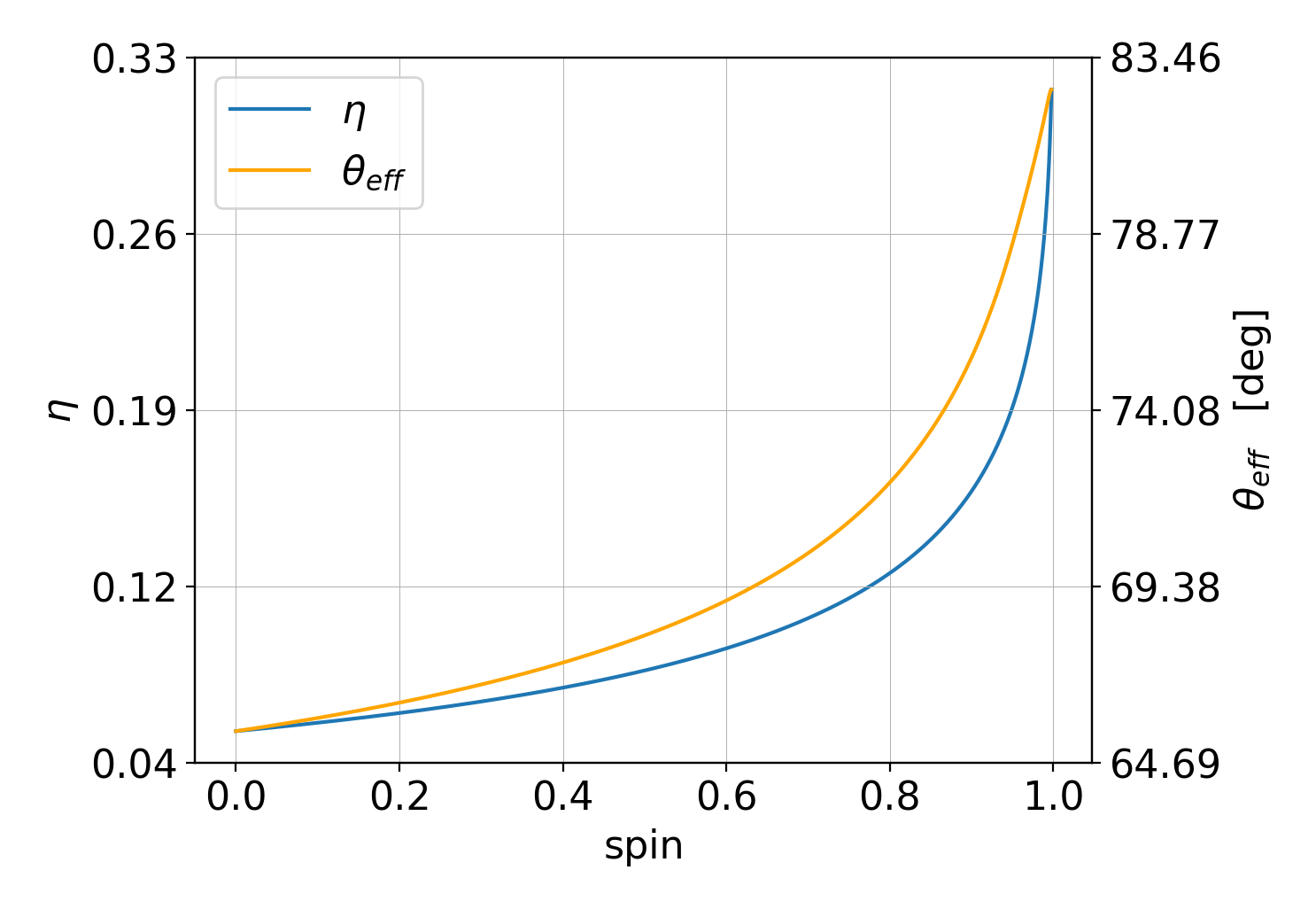}
    \caption{The radiative efficiency (left $y$-axis) and the effective radiation semi-opening angle (right $y$-axis) as a function of the spin parameter $a$.}
    \label{fig:eta}
\end{figure}
In order to quantify the degree of anisotropy as a function of the BH spin, we define an {\it effective semi-opening angle} $\theta _\textrm{eff}(a)$,
 as the truncation angle of the corresponding isotropic emission such that its
 angle-integrated output equals that of the actual angle-dependent emission.
 In Fig.~\ref{fig:eta} we show $\theta_\textrm{eff}(a)$ together with $\eta(a)$.
 From this figure we see that we have a smaller $\theta_\textrm{eff}$, i.e, a more collimated and anisotropic angular pattern, for low spin values, whereas the radiation distribution is more isotropic (larger $\theta_\textrm{eff} \simeq 90^\circ$) for high spin values. 
Similarly, from Fig. \ref{fig: angular_pattern} we see that in spin close to zero, the radiative flux is vertically focused along the spin axis and it decreases with increasing $\theta$. On the other hand, the flux reduction for high $\theta \lesssim 90^\circ$ becomes less pronounced for increasing spin values, i.e., the flux collimation decreases moving towards a nearly isotropic radiation pattern.

\subsection{AGN feedback}\label{AGNfb}

As \cite{Ishibashi19} and \cite{Ishibashi20} pointed out, if the radiation from the accretion disc couples to the surrounding material by exerting radiation pressure on dust or by launching line-driven AGN winds, then the emerging outflow inherits the anisotropy of the impinging radiation, which, in turn, it is shaped by the BH spin. This outflow spin-dependent anisotropy can affect the outflow ability to couple with the ISM and hence it may change the impact that AGN feedback has on the host galaxy and on the MBH growth. In this way, the MBH spin, through its influence on AGN feedback, can possibly play a role in the MBH-galaxy host co-evolution.

Galactic AGN driven outflows in the so-called quasar mode are typically described as the result of the interaction between the disc radiation with the surrounding material and with the transfer of energy and momentum to it. Many works assume that this coupling occurs at the scale of the accretion disc and broad line region (BLR), launching an AGN wind that mediates the interaction between the AGN radiaton and the ISM 
\citep{K03,Zubovas12,CAFG12,Costa14,Hartwig18,Costa20,Torrey20}. Another possibility is that the radiation escapes the inner region and couples at the ISM scale directly through radiation pressure on dust \citep{Murray05,Ishibashi15,Thompson15,Costa18,Barnes18}. In the following we are going to focus on the first mechanism, i.e. AGN winds driven outflows.

\subsubsection{AGN winds}\label{sec:winds}
Observations of bright quasars with blueshifted X–ray absorption lines \citep{Pounds03,Reeves09,Tombesi15} give strong evidence for intense winds from galactic nuclei characterized by relativistic velocities. Such winds are believed to be launched by Thompson scattering or line-driven momentum transfer of AGN radiation to the gas. If each photon emitted by the AGN interacts just once with the gas, then the transferred momentum flux equals the radiation momentum flux $L/c$. If the gas is dusty, then
the optical and ultraviolet (UV) radiation is absorbed and re-emitted at infrared (IR) wavelengths. If the gas is optically thick in the IR, instead of streaming out, the reprocessed IR photons undergo multiple scatterings. In this case, the net momentum imparted by the AGN radiation field may exceed $L/c$. More generally we indicate with $\tau L/c$ the momentum transferred to the gas, where the parameter $\tau$ holds the information concerning the radiation-gas coupling and $\tau = 1$ ($ > 1$) in the single(multi)-scattering scenario.
Then, the total momentum carried by the wind can be written as
\begin{equation}
    \dot{P}_\textrm{w} = \tau \frac{L}{c}.
    \label{Eq: mom conservation}
\end{equation} 
Using Eq. (\ref{Eq: normalization}) and by writing the momentum loading as $\dot{P}_\textrm{w} \equiv \int \dot{p}_\textrm{w}(\theta)r^2\,d\Omega$, where $\dot{p}_\textrm{w}(\theta)$ is the wind momentum flux in the direction $\theta$ at a distance $r$ from the AGN, Eq. (\ref{Eq: mom conservation}) yields
\begin{equation}
    \dot{p}_\textrm{w}(\theta; a) = \tau \frac{L}{c}\frac{f(\theta; a)}{4\pi r^2}.
    \label{eq: wind momentum}
\end{equation}
That is, the momentum flux of the wind follows the same angular pattern of the radiation. Assuming that the wind is launched at a constant velocity $v_\textrm{w}$, then the mass flux in the direction $\theta$  
is simply $\dot{m}_\textrm{w}(\theta; a) = \dot{p}_\textrm{w}(\theta; a)/v_\textrm{w}$, and hence the mass flux angular distribution of the wind follows the same angular pattern as well. By combining Equations \eqref{Eq:eta} and \eqref{Eq: mom conservation} instead we obtain the total mass loading:
\begin{equation}
    \dot{M}_\textrm{w} \equiv \frac{\dot{P}_\textrm{w}}{v_\textrm{w}} = \eta _\textrm{w} \dot{M}_\textrm{acc}.
    \label{Eq: mw-macc}
\end{equation}
where we have defined the mass loading factor
\begin{equation}
  \eta_\textrm{w} = \eta(a) \tau \frac{c}{v_\textrm{w}}.
  \label{Eq: eta wind}
\end{equation}

\subsubsection{Wind-driven outflows}\label{Ed-Md}
The further evolution of wind-driven outflows has been discussed in many papers \citep{Weaver77,Koo92,K03,Costa14,Hartwig18} and textbooks \citep{Dyson97}, in the context of both stellar and AGN feedback. We briefly review some aspects here. 

Once launched, the AGN wind moves in the ambient medium shovelling the material it encounters along its path, forming a shell expanding at constant velocity $\sim v_\textrm{w}$. This phase is referred to as free-expansion and lasts approximately until the swept-up mass equals the mass of the impinging wind. If the ambient medium has uniform density $\rho_0$, the free expansion timescale reads
\begin{equation}
    t_\textrm{free} = \Biggl(\frac{3}{4\pi}\Biggr)^{1/2} \Biggl(\frac{\dot{M}_\textrm{w}}{\rho_0 v_\textrm{w}^3}\Biggr)^{1/2}.
\end{equation}
When the shell radius reaches a distance $\sim R_\textrm{free} \equiv v_\textrm{w}t_\textrm{free}$, the momentum of the material added to the shell start causing it to slow down significantly and free-streaming brakes down. From now on, the incoming wind forms a strong reverse shock against the slowing down shell and a significant fraction of the wind kinetic energy is thermalised.
We can now distinguish a four-layers structure formed by the AGN wind, the shocked wind, the shocked ambient medium and finally the unperturbed eniviornment. The dynamics of this structured shell, or outflow, has been studied extensively both theoretically \citep{Weaver77,Koo92, Dyson97,CAFG12,Zubovas12, King15} and numerically \citep{Costa14,Costa20}. In a nutshell, the subsequent evolution of the outflow depends on the ability of the shocked wind to preserve its thermal energy, which in turn depends on the cooling processes involved and on the associated timescales. If radiative losses in the shocked wind are negligible, it expands adiabatically doing `$PdV$' work on the shocked ambient medium, driving an ``energy driven'' outflow \citep{King05}. If the shocked wind is radiatively cooled, the shell of swept-up ambient gas is driven solely by the wind's ram pressure \citep{K03}. Such solutions are termed ``momentum driven''. If the shocked wind does cool, but inefficiently, the wind solution is intermediate between momentum and energy-driven \citep{CAFG12}. 

From an observational point of view, both energy and momentum driven outflows have been detected \citep{Fluetsch19,Tozzi21}. However, the observed outflows are mostly on kpc-scales, while theoretical arguments \citep{CAFG12,King15} and the lack of observational evidence of cooling wind
shocks \citep{Bourne13} suggest that under realistic circumstances the shocked wind bubble cools inefficiently, and that momentum-driven outflows should be confined within the central few 100 pc from the MBH. Therefore, outflows observed in the momentum driven regime are more consistent with a radiation pressure on dusty gas mechanism or with an energy-driven outflow poorly coupled with the ISM, rather than with an efficiently cooling wind-driven outflow. 

\subsubsection{Energy \& momentum outflows driven by anisotropic winds}

We derive now the analytic solution for the propagation in an homogeneous medium of an outflow driven by an anisotropic wind, both in the energy- and momentum-driven scenarios. 
\begin{figure}
    \centering
    \includegraphics[scale=0.7]{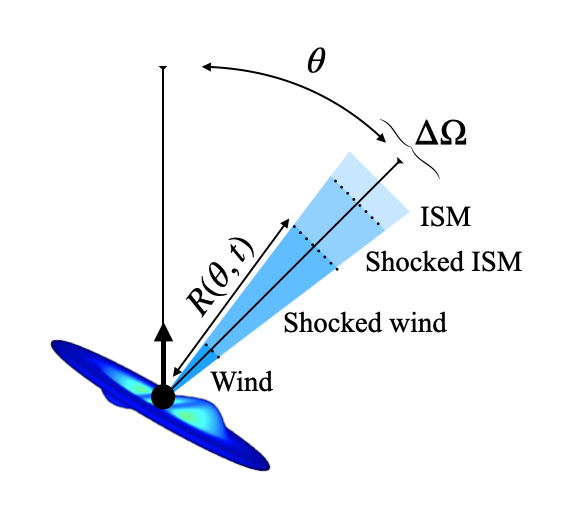}
    \caption{Schematic view of the outflow structure in a slice subtended by a solid angle $\Delta \Omega$ and centered in a direction forming an angle $\theta$ with the MBH spin. This stratified structure comprises the AGN wind, the shocked wind (which extends up to a distance $R(\theta,t)$ from the MBH), the shocked ISM, and the unperturbed ISM.  The illustration also shows the warped accretion disc that feeds the spinning MBH.}
    \label{fig:model}
\end{figure}

We characterize the evolution of the outflow by calculating the location $R(\theta,t)$ of the contact discontinuity that separates the shocked wind from the shocked ambient medium (see Fig. \ref{fig:model}). In the energy driven regime, the shocked wind shell is hot and thick and its thermal energy evolution is due to the energy injected by the wind (suddenly converted in thermal energy) and the work done on the above shocked ambient medium. Using Equations \eqref{Eq:eta} and \eqref{Eq: mw-macc}, we have that the thermal energy is added to the shock wind layer at a rate $1/2\dot{M}_\textrm{w}v_\textrm{w}^2 = 1/2 \tau (v_\textrm{w}/c) L \equiv \epsilon L$, where $\epsilon = 1/2 \tau v_\textrm{w}/c$.
Now, if we consider a single slice of the outflow, in the direction $\theta$ and subtended by a solid angle $\Delta \Omega$, as in Fig. \ref{fig:model}, then the impinging disc luminosity in the slice direction is $Lf(\theta,a)\Delta \Omega /4\pi$  (see Eq. \ref{Eq: normalization}) and a fraction 
$\epsilon$ of it is converted in thermal energy in the shocked wind layer of the slice. Then, if $P$ is the pressure of the shocked wind layer, the pressure force exerted on the layer above can be written as $P\Delta \Omega R^2$ and the $PdV$ work
 as $P \Delta \Omega R^2 dR$. Assuming that
the shocked wind layer is thick enough to neglect the portion of the slice occupied by the freely streaming wind, its volume can be approximated with that of the slice up to $R$, i.e. $\Delta \Omega R^3 /3$, and hence its internal energy with $(3/2)P\Delta \Omega R^3 /3$. Then, if $\rho _0$ the ambient medium density we can write the conservation of the shocked wind energy and the conservation of the shocked ambient medium momentum of the slice as
\begin{align}
    & \frac{d}{dt}\frac{3}{2}\Biggl( \frac{\Delta \Omega}{3} R^3 P\Biggr) = \epsilon Lf(\theta;a)\frac{\Delta\Omega}{4\pi} - \Delta \Omega R^2 \dot{R} P, \label{Eq:energy shocked wind}\\
    &\frac{d}{dt}\Biggl(\rho_0 \frac{\Delta \Omega}{3} R^3 \dot{R}\Biggr) = R^2 \Delta \Omega P, \label{Eq:momentum shocked ambient medium }
\end{align}
respectively. By solving Eq.~(\ref{Eq:momentum shocked ambient medium }) for $P$ and replacing in Eq.~(\ref{Eq:energy shocked wind}) we get 
\begin{equation}
    10 R^2 \dot{R}^3 + 8 R^3\dot{R}\ddot{R} + \frac{2}{3}R^4 \dddot{R} = \frac{v_\textrm{w}Lf(\theta; a)}{2 c \pi \rho_0},
\end{equation}
which admits the self-similar solution 
\begin{equation}
    R_\textrm{sh}(\theta, t) = \Biggl(\frac{125}{308 \rho_0\pi}\frac{v_\textrm{w}}{c} Lf(\theta; a) \Biggr)^{1/5}t^{3/5}.
    \label{Eq: shell energy driven}
\end{equation}
Eq.~\eqref{Eq: shell energy driven} describes the evolution of the contact discontinuity in time, for each direction $\theta$, in the energy driven regime.

In momentum-driven outflows, the energy of the
shocked wind is quickly dissipated on small scales via radiation losses and the shocked wind shell collapses into a thin layer, as its thermal pressure support is radiated away. Then, the shocked ambient medium layer is driven outward directly by the ram pressure of the impinging AGN wind, which equals the momentum flux of the disc radiation if $\tau = 1$. 
In this case, we write the momentum conservation of the shocked ambient medium in the slice as  
\begin{equation}
    \frac{d}{dt}\Biggl( \rho_0 \frac{\Delta \Omega R^3}{3} \dot{R} \Biggr) = \frac{Lf(\theta;a)\Delta \Omega}{4\pi c},
\end{equation}
which is solved by 
\begin{equation}
    R_\textrm{sh}(r,\theta,t) = \Biggl( \frac{3Lf(\theta;a)}{2\pi c \rho_0}\Biggr)^{1/4}t^{1/2}.
    \label{Eq: shell momentum driven}
\end{equation}
Eq.~\eqref{Eq: shell momentum driven} represents the evolution of the contact discontinuity in the momentum driven regime. 
We briefly mention that the main cooling process that could make this regime possible is the Compton cooling of free electrons in the shocked wind shell against the AGN photons, as discussed in \cite{K03,CAFG12, Hartwig18,Richings18, Costa20, Torrey20}. Following \cite{Sazonov04}, if we assume that the AGN radiation field has a nearly obscuration-independent Compton temperature $T_\textrm{AGN} = 2 \cdot 10^7$ K, the gas Compton heating/cooling rate for gas with temperatures $T < 10^9$ K (non relativistic regime) is 
\begin{equation}
    \Lambda _\textrm{Cpt} = \frac{n_e \sigma_\textrm{T}L}{m_ec^2 \pi r^2} k_\textrm{B}(T-T_\textrm{AGN}),
    \label{Eq: LambdaCpt}
\end{equation}
where $T$ is the gas temperature, $m_e$ the electron mass, $\sigma_\textrm{T}$ the Thompson cross-section, $n_e$ the electron number density (bound + free) and $r$ the distance of the gas element from the AGN.

\section{Feedback implementation}\label{sec:model}

In order to investigate the role of anisotropic radiative feedback in more realistic scenarios, we implemented the anisotropic spin-dependent AGN wind discussed above (section \ref{sec:winds}) in the code \textsc{gizmo} \citep{Hopkins15}. In a nutshell, our implementation is characterized by a sub-grid accretion disc whose properties (mass $M_\alpha$ and angular momentum $\mathbf{J}_\alpha$) evolve due to the accretion $\dot{M}_\textrm{in}$ from resolved scales on the sub-grid disc (modelled with a modified Bondi-Hoyle prescription similar to \citealt{Tremmel16}) and due to the accretion $\dot{M}_\textrm{acc}$ within the sub-grid disc on the MBH. Similarly, the physical (sub-grid) MBH is characterized by its mass $M_\bullet$ and its angular momentum $\mathbf{J}_\bullet$ and they evolve according to the sub-grid accretion. This model, based on \cite{Cenci21} and \cite{Sala21} (see also \citealt{Fiacconi18}), tracks the exchange of angular momentum between the MBH and the sub-grid disc due to both the accretion on the MBH and the Bardeen-Petterson torque \citep{BP75}, in this way following the MBH spin evolution consistently with evolution of the sub-grid disc properties. More details about the accretion and spin evolution models are provided in Appendix \ref{App: acc implementation}. 

In the following we present our AGN feedback model, based on the wind-spawning technique developed by \cite{Torrey20}, whose approach is similar to that devised by \cite{Costa20} for \textsc{arepo}. Contrary to the AGN feedback models based on thermal/kinetic energy injection in the MBH neighbours gas particles, \cite{Torrey20} and \cite{Costa20} directly inject the wind mass, momentum and energy from a sub-grid region into the resolved scales and let the hydrodynamical evolution make the wind shock and thermalise. 
While in these models the wind injection is assumed isotropic, here we force the wind momentum (Eq. \ref{eq: wind momentum}) and mass fluxes to follow the sub-grid disc luminosity angular pattern resulting from gravitational bending, which is ultimately set by the MBH spin. We mention that another source of AGN feedback anisotropy does exist when AGN winds are accelerated at relativistic velocities. Indeed, an isotropic AGN radiation source driving a nuclear wind with relativistic velocity is perceived as anisotropic by the wind gas itself due relativistic beaming \citep{Luminari20}. However, this effect translates in angular variations of $v_\textrm{w}$ of a few percent for $v_\textrm{w} = 0.01c$, as we will assume later, which is negligible compared to the variation in the wind momentum attributed to the photons gravitational bending.  

Overall, the MBH mass and spin are evolved in a sub-grid fashion according to the properties of the sub-grid accretion disc, which are influenced by the resolved accretion flow on the sub-grid system. The presence and characteristics of such an inflow are affected by the feedback magnitude and anisotropy, which in turn are set by the sub-grid disc and MBH properties. This results in a complex non-linear interplay between MBH fueling, feedback, its anisotropy and the MBH spin, which our sub-grid model is aimed to capture. 
In the following we provide more details about the wind launching implementation.

\subsection{AGN wind injection}\label{Sec: agn wind}
Following \cite{Torrey20} and \cite{Costa20}, we assume that a factor $\tau$ of the radiation momentum flux is transferred to the gas at sub-grid scales, generating an AGN wind which is directly injected into the resolved scales. 
The wind mass outflow rate $\dot{M}_\textrm{w}$ is computed from the unresolved disc accretion rate as shown in Eq. (\ref{Eq: mw-macc}), where both $v_\textrm{w}$ and $\tau$, appearing in the definition of $\eta_\textrm{w}$ (Eq. \ref{Eq: eta wind}), are free parameters of the model. In practice, the wind is simulated by removing mass from the sub-grid disc and by spawning $N_\textrm{w}$ new gas particles at a rate $\dot{M}_\textrm{w}$. The newborn wind particles are distributed uniformly on a sphere centered on the MBH and with radius equal to the minimum between one tenth of the MBH gravitational softening $\epsilon_\bullet$ and half of the smallest MBH-gas particle separation. The mass distribution of the spawned particles follows the sub-grid disc luminosity angular pattern. More precisely, the mass of the $i-$th spawned particle, with polar angle $\theta _i$ from the MBH spin direction, is assigned as 
\begin{equation}
    m_i = \frac{ f(\theta_i, a)}{\sum_j f(\theta_j, a)}M_\textrm{spawn},
\end{equation}
where the sum at denominator spans over all the spawned particles, $M_\textrm{spwn}$ is the total mass to be spawned and $f(\theta,a)$ is the luminosity angular pattern defined in Eq. (\ref{Eq: normalization})

The wind particles are then launched radially outward with constant velocity $v_\textrm{w}$ and by interacting with the other surrounding gas particles they generate an outflow. During their evolution, the wind particles can be merged into non-wind gas particles, properly transferring to them their mass, momentum, and energy as in an inelastic collision, see Appendix E of \cite{Hopkins15}. 
We required that such a merger occurs once the velocity of the target wind particle falls below five times the mean velocity of the non-wind particles in its kernel. 
In order to track the propagation of the wind into the environment, we introduced a scalar wind tracer $\zeta$, similarly to \cite{Costa20},
that represents the wind mass fraction of a gas particle. Therefore $\zeta = 1$ for wind particles, non-wind particles are initialized with $\zeta = 0$ and $0< \zeta < 1$ values characterize non-wind particles that experienced mergers with wind particles.

In order to be able to capture the formation of a momentum-driven outflow our model includes gas Compton cooling. We assume that the AGN photons interact with the surrounding gas at sub-grid scale, launching the AGN wind, and then they are remitted isotropically. Then, such reprocessed photons can 
scatter with the high energy electrons of the shocked wind layer of the outflow, making it loose energy and cool. In practice, following \citep{Hopkins16}, a contribution $\Lambda _\textrm{Cpt}$ as in Eq. (\ref{Eq: LambdaCpt}) is included when computing radiative cooling/heating of gas particles, where now $r$ is the distance between the gas particle and the MBH and $L$ is the instantaneous sub-grid disc luminosity computed from Eq. (\ref{Eq:eta}).

In the sub-grid model described in the present section, the MBH timestep $dt_\bullet$ is taken small enough to resolve the sub-grid accretion, wind launching and spin evolution and large enough to guarantee that the disc attains a steady-state warped profile, as assumed in our prescriptions \citep{Cenci21}. 

\section{Tests}\label{Sec: tests}
\begin{table*}
  \centering
    \small{\begin{tabular}{c cccccccc}
    \hline\hline\vspace{-0.75em}\\
        & $a$ & $\eta$ & $\log (M_\bullet/M_\odot)$ & $f_{\rm Edd}$ & $L\rm [erg/s] $ & $\mu n\rm [cm^{-3}]$ &  $\epsilon_{\rm grav}\rm [pc]$ & $N_{\rm w}$ \bigstrut\\
    \hline\vspace{-0.75em}\\
    \multirow{3}[2]{*}{Energy driven} & 0.01 & 0.057 & 7 & 0.1  & $1.2\cdot 10^{44}$ & 1 &  0.11 & 588   \\
          & 0.95 & 0.190 & 7    & 0.332 & $3.98\cdot 10^{44}$ &1 &  0.11 &  588 \\
          & 0.9982 & 0.323 & 7 & 0.567 &  $6.8\cdot 10^{44}$ &1 &  0.11 & 588 \bigstrut[b]\\
    \hline\vspace{-0.75em}\\
    \multirow{3}[2]{*}{Momentum driven} & 0.01 & 0.057 & 10 & 0.3 & $3.6 \cdot 10^{47}$ & $10^6$ &  0.08& 108  \bigstrut[t]\\
          & 0.95 & 0.190 & 10    & 0.997 & $1.2 \cdot 10^{48}$ & $10^6$ &  0.08& 108  \\
          & 0.9982 & 0.323 & 10 & 1.7 & $2.04 \cdot 10^{48}$ & $10^6$ &  0.08& 108  \bigstrut[b]\\
    \hline\hline\vspace{-0.75em}\\
    \end{tabular}}%
    \caption{Summary of the parameters adopted in our simulations. In all simulations we employed $10^7$ gas particles and wind particles velocity and temperature are initialized with $v_\textrm{w} = 0.01 c$ and $T_\textrm{w} = 2\cdot 10^4$ K. 
    }
  \label{tab: tests}%
\end{table*}%
\begin{figure*}
    \centering
    \includegraphics[scale=0.5]{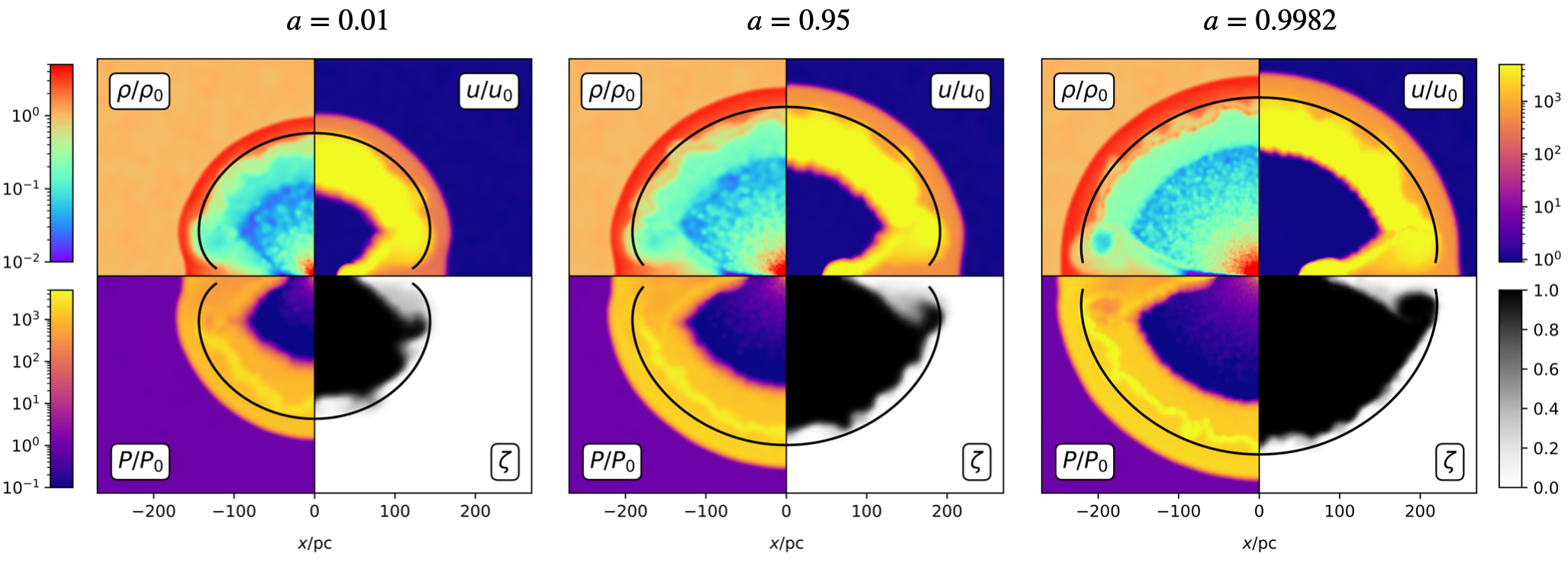}
    \centering
    \includegraphics[scale=0.5]{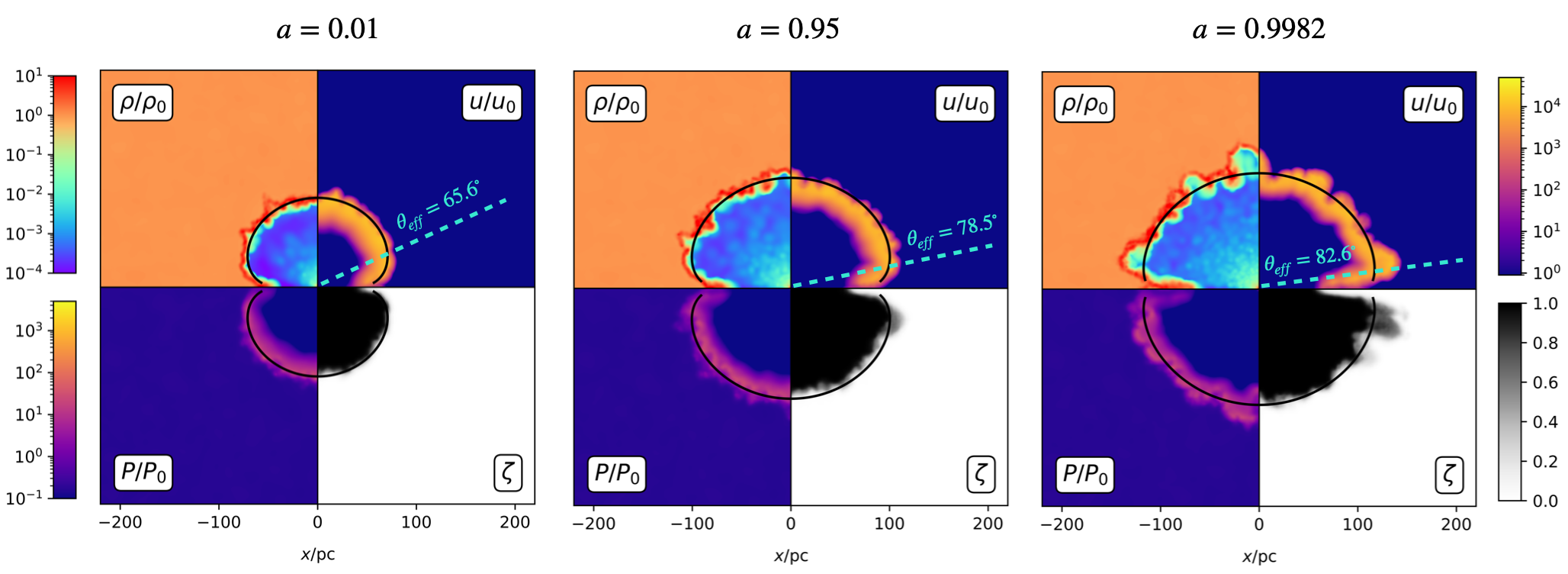}
    \caption{Final snapshots of the propagation of an AGN wind-driven outflow for different spin values (increasing from left to right) and the two considered regimes (energy-driven in the top row and momentum-driven in the bottom one).  Each panel illustrates the density, temperature and pressure fields in units of the corresponding quantities of the assumed background medium as well as the wind mass fraction. 
In the energy-driven simulations the outflow can be divided into four distinct sections: (1) the freely-expanding wind, (2) the shocked wind, (3)
the shocked ambient medium and (4) the undisturbed ambient medium. In the momentum-driven simulations radiative cooling makes the shocked wind layer cool and regions (2) and (3) are condensed into a thin shell. The wind tracer is injected together with the wind and is therefore only present in regions (1) and (2).
The black lines correspond to the analytical location of the contact discontinuity between regions (2) and (3), as computed in Eq. (\ref{Eq: shell energy driven}-\ref{Eq: shell momentum driven}). In momentum-driven simulations the cyan dashed lines indicate
the effective aperture of the radiation 
as defined in \ref{ang pattern}.
}
    \label{fig: tests}
\end{figure*}

In order to compare the simulated evolution of the wind-driven outflows against the analytical predictions given by Eqs. (\ref{Eq: shell energy driven}, \ref{Eq: shell momentum driven}), we first adopt an idealized setup. We consider an active MBH embedded in a gas with uniform density and temperature, so to ensure that the outflow anisotropy is the result of the intrinsic anisotropy of the wind, rather than of the ambient density distribution. 
In order to compare the simulated outflows with Eqs.~(\ref{Eq: shell energy driven},\ref{Eq: shell momentum driven}), we switch off gravitational forces so that the outflow evolution is purely hydrodynamical. For all the duration of the simulations we keep the AGN luminosity and angular pattern constant, i.e., we force the properties of the sub-grid disc and MBH not to vary. 
 
In order to test both energy- and momentum-driven outflows, we consider two different setups. In the first case, 
we assume a gas number density $\mu n = 1$\,cm$^{-3}$, where $\mu$ is the mean-molecular-weight, and a MBH mass $10^7 $ M$_\odot$, while in the second $\mu n = 10^6$\,cm$^{-3}$ and MBH mass $10^{10}$ M$_\odot$. In both cases the gas temperature is set at $T = 2\cdot 10^4$ K.
In order to test different outflow anisotropies, for each setup we consider three different values of the MBH spin, $a_\bullet = \{ 0.01, 0.95, 0.9982\}$. 
In the energy-driven simulation with $a_\bullet = 0.01$, we assume $f_\textrm{Edd} \equiv \dot{M}_\textrm{acc}/\dot{M}_\textrm{Edd} = 0.1$ (where $\dot{M}_\textrm{Edd}$ is the Eddington accretion rate), while $f_\textrm{Edd} = 0.3$ in the corresponding momentum-driven simulation. In the simulations with higher MBH spin we keep the same sub-grid disc properties as for $a_\bullet = 0.01$, therefore the corresponding $f_\textrm{Edd}$ values scale as $\eta(a_\bullet)/\eta(a_\bullet = 0.01)$, according to Eq. (\ref{Eq: fedd}). 
Energy-driven simulations run up to $t_\textrm{end} \sim 0.22$ Myr, while momentum-driven ones for $t_\textrm{end} \sim 1.5 $ Myr. 
The values of the main parameters of all runs are summarized in Table \ref{tab: tests}.

Figure \ref{fig: tests} shows the final snapshots for all six simulations, with energy-driven runs in the top panels and momentum-driven in the bottom ones. For each snapshot, the figure reports
 the density, temperature and pressure fields in units of the corresponding quantities of the background medium as well as well as the wind mass fraction.
Superimposed to all maps, we draw as a black curve the location $R_\textrm{sh}(t_\textrm{end},\theta)$ of the contact discontinuity, as computed via Eqs. (\ref{Eq: shell energy driven},\ref{Eq: shell momentum driven}). 
In the energy-driven simulations (top panels) the typical stratified structure of outflows can be recognized, with the free propagation wind layer, the shock wind, the shocked ambient medium and the unperturbed ambient medium. The simulated location of the contact discontinuity approximates very well its theoretical prediction (Eq. \ref{Eq: shell energy driven}). 
In the momentum-driven simulations (bottom panels) Compton cooling makes the shocked wind layer cool and collapse,  forming a thin shell that separates the free expanding wind from the unperturbed medium. Again, the location of the separating shell is in good agreement with the theoretical estimate (Eq. \ref{Eq: shell momentum driven}).

In both energy- and momentum-driven simulations, as the spin increases, the angular pattern of the emerging outflow becomes more spherical and spreads at larger distance from the MBH. This is expected, as larger spin values yield more isotropic luminosity angular patterns as well as more luminous discs.

\section{Isolated galaxy simulations}\label{sec: isolated galaxy}

\begin{figure*}
    \centering
    \includegraphics[scale=0.55]{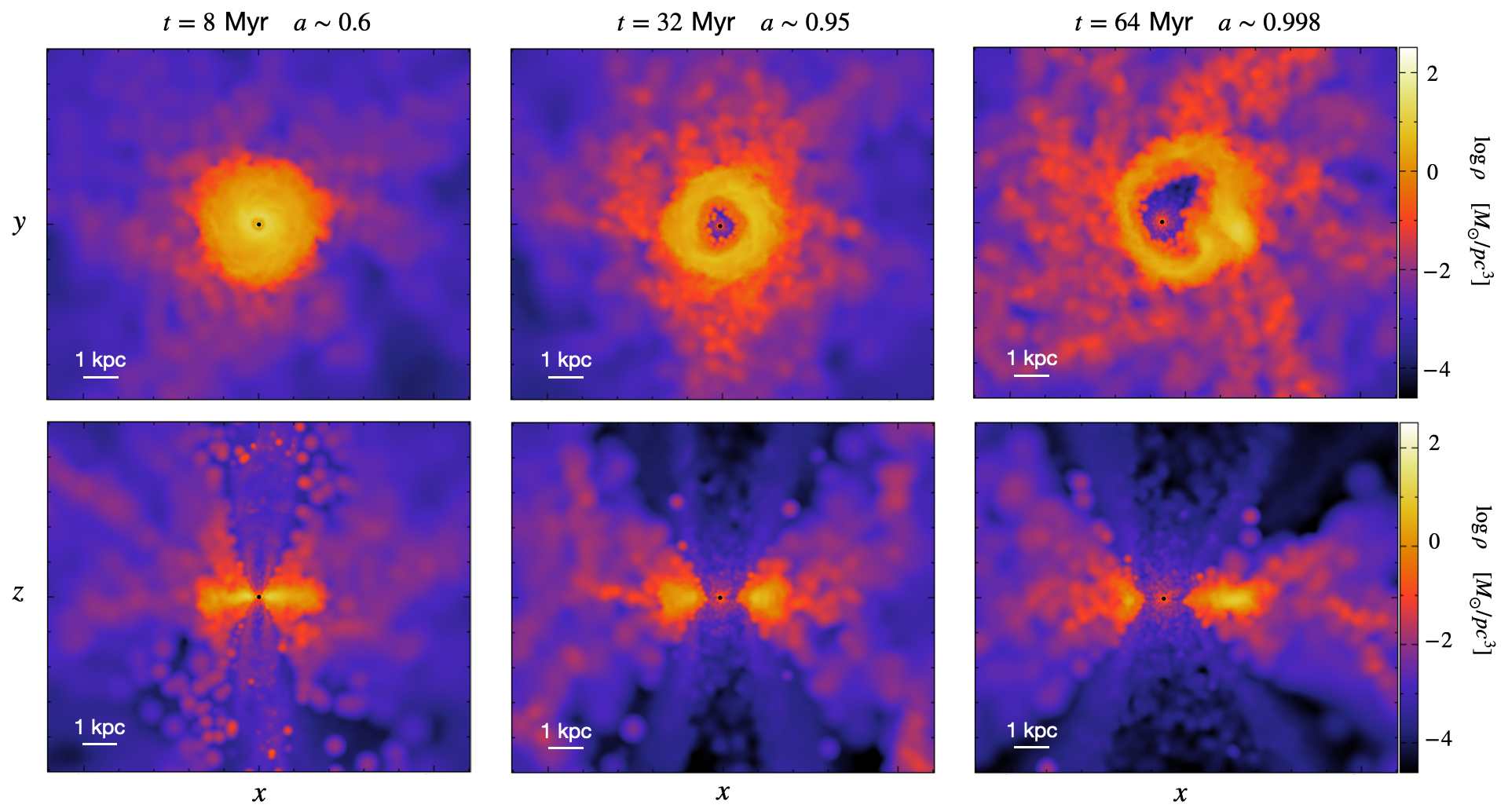}
    \caption{ Density slices 
    of three snapshots from simulation {\labsim{C\_f}}, with the disc face-on  in the top panels and side-on in the bottom panels. The MBH location is marked by a black dot. }
    \label{fig: C_f}
    \centering
    \includegraphics[scale=0.55]{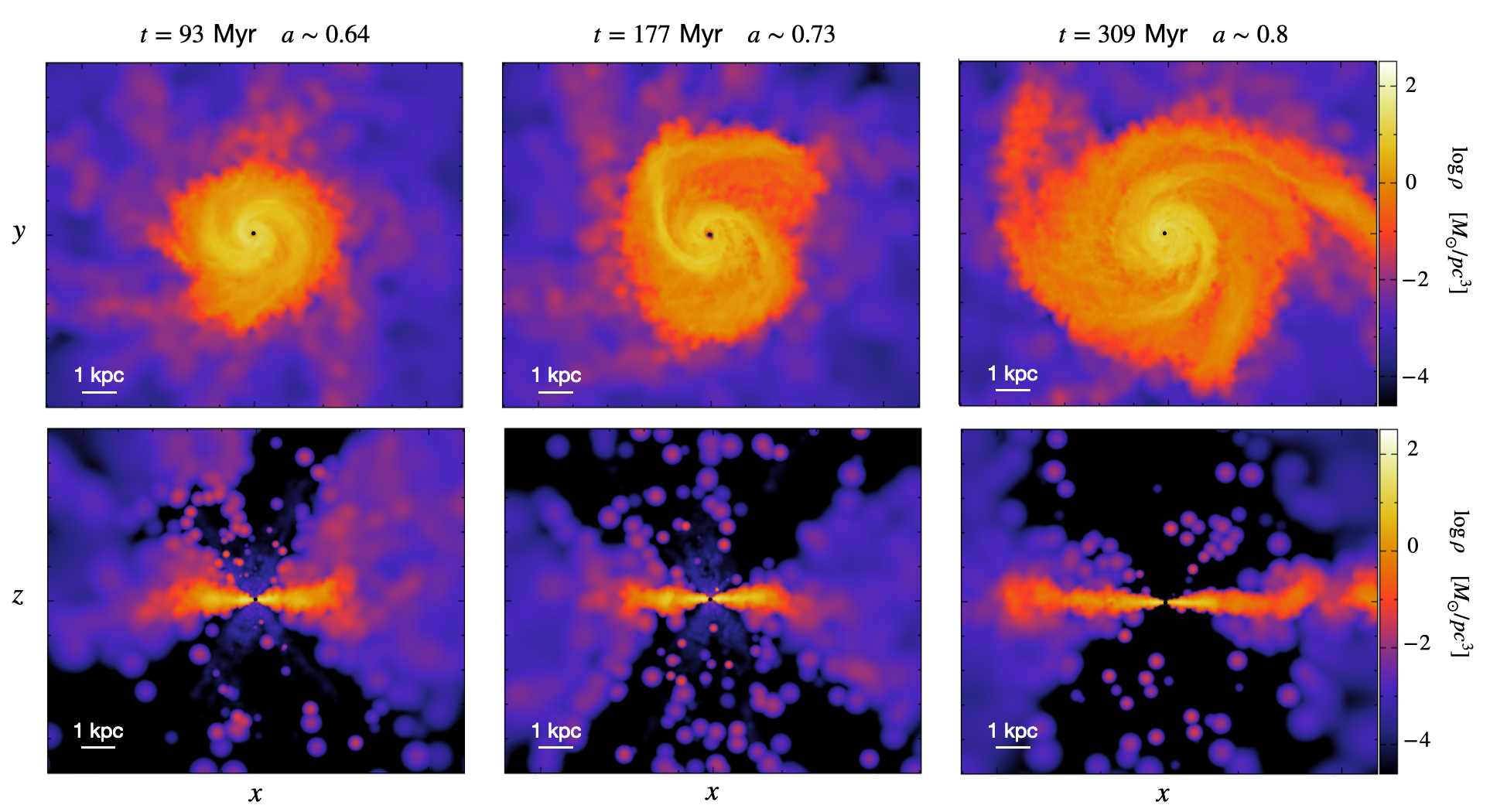}
    \caption{Density slices 
    of three snapshots from simulation {\labsim{E\_f}}, with the disc face-on  in the top panels and side-on in the bottom panels. The MBH location is marked by a black dot.}
    \label{fig: E_f}
\end{figure*}

\begin{figure*}
    \centering
    \includegraphics[scale=0.57]{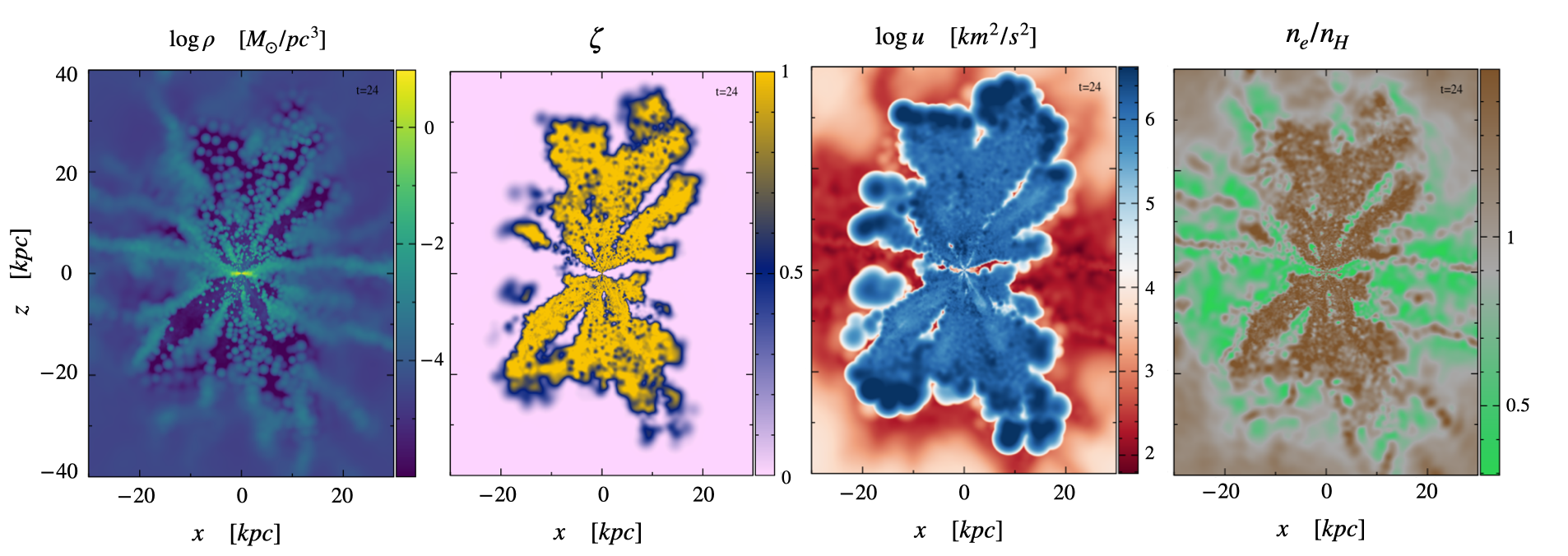}
    \caption{ From left to right, edge-on maps of the gas density, the wind mass fraction, the internal energy, and the electron abundance fraction of simulation {\labsim{E\_f}} at $t=24$ Myr.}
    \label{fig: Warhole}
\end{figure*}

We employ now the sub-grid model for spin-dependent AGN winds described in Section \ref{sec:model} to study the role of wind anisotropy in shaping AGN-driven galactic outflows, and then the impact such outflows have on the host galaxy. We start by constructing an isolated galaxy setup similarly to \cite{Costa20}. We first initialize a gaseous halo of mass $M_{g} = f_{g}M_{200}$, where $M_{200} = 10^{12} M_\odot$ is the halo (dark matter + gas) virial mass and $f_g = 0.17$ the gas mass fraction, with a Navarro-Frenk-White \citep{NFW97} density profile. A $M_\bullet = 10^8 M_\odot$ MBH is placed at the center of the system and the gas internal energy is set to guarantee hydrostatic equilibrium. The halo concentration parameter is $\mathcal{C} = 7.2$, which 
gives a halo scale radius $a_\textrm{halo} = 28.54$ kpc and a virial radius $r_{200} = \mathcal{C}a_\textrm{halo} = 205.5$ kpc.
The gas initial specific angular momentum follows a profile
$j(r,\theta)$, where $r$ and $\theta$ are the spherical radial and polar coordinates, as in Eqs. (12-15) of \cite{Liao17}. Such profile is characterized by the halo spin parameter $\lambda _\textrm{sim}$, introduced by \cite{Bullock01} (their Eq. 5), which is set to $\lambda_\textrm{sim} = 0.035$. 

Radiative cooling is modelled as described in \citep{Hopkins18}, assuming initial metallicity $0.1 Z_\odot$ and ignoring radiative cooling processes below $T<10^4$ K. Due to the loss of pressure support following radiative cooling, at $t=0$ the spinning gaseous halo starts to collapse towards the center and the inner region settles into a galactic disc continuously fed by the outermost falling layers.

The gaseous halo is sampled with $N = 3\cdot 10^5$ particles only up to a radius of $0.6 \, r_{200}$, as gas particles above this radius take more than $\sim 0.7 $Gyr to fall within $\sim 5$ kpc from the center, where the galaxy forms, and therefore are not relevant for the spatial and temporal scales we consider.  The mass resolution of our simulations is $m_\textrm{gas} = 5.67\cdot 10^5 M_\odot$. Dark matter is not sampled and enters the simulation as a static analytic potential.  

Star formation is treated following \cite{Springel03}, where
the effects of unresolved physical processes operating within
the interstellar medium (ISM) are captured by an effective equation of state that is applied to all gas with hydrogen number density $n_\textrm{H} > n_\textrm{th}$. This
effective equation of state is stiffer than that of isothermal
gas, because it accounts for additional pressure provided by
supernova explosions within the ISM. Stellar particles are spawned stochastically from gas with $n_\textrm{H} > n_\textrm{th}$ at a rate
\begin{equation}
    \frac{d \rho_\star}{dt } = (1-\beta)\frac{\rho_\textrm{c}}{t_\star},
\end{equation}
where $\beta = 0.1$ is the mass fraction of massive stars assumed to instantly explode as supernovae, $\rho_\textrm{c}$ is the density of cold
clouds \citep[see][for details]{Springel03} and $t_\star = t_{\star,0}(n_\textrm{H}/n_\textrm{th})^{-1/2}$ 
is the star formation timescale, with $t_{\star,0} = 1.5$ Gyr and $n_\textrm{th} = 0.5 \text{cm}^{-3}$. Supernova-driven winds are not modelled, as they would add a further layer of complexity affecting both star formation, the MBH growth and hence AGN feedback \citep[see, e.g.,][]{Dubois15}. Therefore, our simulations should be regarded as idealised experiments aimed at illustrating how AGN wind modelling affects the impact of MBH feedback on the host galaxy and on the MBH evolution, without ``contamination'' from supernova-driven winds.

After the first 150 Myr, once a galactic disc has formed and star formation has already reached its peak, MBH accretion and feedback are ``switched on''. We denote this instant with $t_0$. Then, we perform six different simulations
divided in two sets (labelled with {\labsim{C}} and {\labsim{E}}) corresponding to different MBH accretion prescriptions, each set consisting of three different simulations characterized by different feedback anisotropies (indicated with labels {\labsim{f}}, {\labsim{iso}}, {\labsim{a0}}). In the simulation set labelled with {\labsim{C}}, the sub-grid accretion disc is forced to maintain the same properties for the entire duration of the simulations, i.e., the disc mass and angular momentum do not change because of accretion onto the MBH and wind ejection. In these simulations, the Eddington factor $f_\textrm{Edd} \equiv \dot{M}_\textrm{acc}/\dot{M}_\textrm{Edd}$ is kept constant equal to 1  (similarly to  \citealt{Torrey20}, \citealt{Costa20} and \citealt{Mercedes23}), corresponding to a sub-grid disc of constant luminosity $L = 1.2 \cdot 10^{46}$ erg/s.
On the contrary, in {\labsim{E}} simulations, $f_\textrm{Edd}$ is allowed to evolve and its value is set by the instantaneous MBH + subgrid disc properties (Eq.~\ref{Eq: fedd}), which, in turn, are determined by the inflow on the disc, the accretion on the MBH and wind ejection (see Eqs. ~\ref{Eq: MBHdot}-\ref{Eq: dotJalpha}). In this way, {\labsim{E}}-simulations are able to capture the mutual non-linear influence that inflows and outflows have on each other, that lead to a self-regulated MBH growth.\footnote{\cite{Torrey20} showed that an Eddington accreting AGN generates wide cavities in its surroundings, but noted that this lower density should reflect in decreased inflow on the AGN in turn reducing his power and the cavity size itself, letting the system self-regulate.} In addition to these simulation sets, we prolonged the simulation for the spinning halo collapse beyond 150 Myr without switching on the AGN, so that the sets {\labsim{C}} and {\labsim{E}} can be compared to the case in which the MBH is not active. We indicate this simulation with {\labsim{NoFb}}.

Both {\labsim{C}} and {\labsim{E}} sets consist of three simulations, accounting for a different wind anisotropy. Simulations {\labsim{f}} have an angular pattern $f(\theta;a)$, as illustrated in section~\ref{ang pattern}, that evolves according to the evolution of the spin parameter $a$; {\labsim{iso}}-simulations are characterized by an isotropic angular pattern, i.e. $f=1$; {\labsim{a0}}-simulations assume instead a fixed angular pattern corresponding to the one of $a \simeq 0$, $f=f(\theta; 0.01)$.Note that, independent of the anisotropy pattern, the MBH spin and $\eta(a)$ are allowed to evolve in these runs. The motivation for these choices of angular patterns is the following: real outflows anisotropy can result from the wind ``intrinsic'' anisotropy imparted by the radiation angular pattern and linked to the MBH spin, i.e. the anisotropy discussed in section~\ref{ang pattern} and seen in Fig.~\ref{fig: tests}, or it can be induced by the anisotropy of the medium in which the outflow propagates as, for example, in a spiral galaxy where the gas density is higher in the midplane of the galaxy than perpendicular to it. In this way, an ``intrinsically'' spherical outflow expands more easily along the galaxy axis, i.e., along the least resistance path, turning into a bipolar (anisotropic) outflow \citep{Hartwig18, Costa20, Zubovas23}.\footnote{This may explain, for example, the presence of the Fermi bubbles in the Milky Way \citep{KZubovas12}.} Since outflow anisotropy is shaped by these two factors, both present in simulations {\labsim{f}}, a comparison with simulations {\labsim{iso}} allows to disentangle these effects and to properly assess the relevance of radiation pattern anisotropy in galactic outflows. Instead, simulations {\labsim{a0}} represent the opposite term of comparison, in which the angular pattern is maximally anisotropic, i.e. $f(\theta; 0.01) \simeq 2\cos \theta$ as in (Newtonian) $\alpha$-discs, as if all the isotropy coming from gravitational bending were removed. 
 
In both {\labsim{C}} and {\labsim{E}} sets, we start the simulations at $t_0$ with $f_{\textrm{Edd}} = 1$, $a = 0.01$, $M_\alpha = 0.005 M_\bullet$ and $M_\bullet + M_\alpha = 10^8 M_\odot$. The wind velocity and temperature are fixed to $v_\textrm{w} = 0.01\,c$ and $T_\textrm{w} = 2\cdot 10^4$. $N_\textrm{w} = 48$ wind particles are spawned at each spawning event and the radiation-wind coupling parameter is fixed to $\tau = 1$. In all simulations the gravitational softening for all particle species is 2 pc.

In the following sub-sections, we will discuss the impact of AGN feedback in our setups, focusing on the one hand on its effect on the host galaxy, in particular on the central gas reservoir and star formation, and on the other hand on its influence on the MBH evolution in mass and spin.
Before entering in details, we give a qualitative glimpse on how the galaxy-MBH evolution looks like in simulations {\labsim{C\_f}} and {\labsim{E\_f}}. 

Figure \ref{fig: C_f} shows three snapshots of the simulation {\labsim{C\_f}}, both face on (top panels) and side-on (bottom panels). In the side-on view, we see that after $t \simeq 8$ Myr from the AGN ``switch on'', a collimated wind is piercing the circum-galactic-medium (CGM), its opening angle widening over time. Correspondingly, in the face-on view, a $\sim$kpc-scale cavity is cleared in the central region surrounding the MBH, becoming increasingly larger with time. 
Both these effects are accompanied by an increase of the spin value, which reaches $\sim 0.998$ after $\simeq 64$ Myr, and hence by an increase in the disc radiative efficiency and in the isotropy of the radiation angular pattern (see discussion in section~\ref{ang pattern}).  We remark that in this simulation the accretion rate on the MBH is forced to be constant, equal to the Eddington rate. However, as noted by \cite{Torrey20}, once the gas reservoir feeding the MBH diminishes with the formation of a central cavity, the accretion on the MBH should decrease as well, reducing the AGN power and hence its ability to further enlarge the cavity. We show this effect in Fig. \ref{fig: E_f}, which is the analogous of Fig. \ref{fig: C_f} for the simulation {\labsim{E\_f}}, in which the MBH accretion is consistently evolved with the inflow on the disc and the wind ejection from it. In this case, no cavity formation can be seen in the face-on view and a milder wind (compared to Fig. \ref{fig: C_f}) is seen in the side-on view. This suggests  that a much lower accretion (and hence outflow) rate is achieved once that the disc is allowed to adjust itself through its interaction with the environment.
In particular, we will see in section~\ref{sec: fedd} that, in this simulation, $f_\textrm{Edd}$ indeed drops by $\sim 1\div 2$ orders of magnitude from its initial value.

In Figure \ref{fig: Warhole} we show a zoom-out of the side-on view of simulation {\labsim{E\_f}} at $t = 24$ Myr. The four columns illustrate, from left to right, the gas density, the wind mass fraction, the gas internal energy and the electron abundance fraction $n_e/n_\textrm{H}$. Even though the AGN wind doesn't seem to affect much the host galaxy (Fig. \ref{fig: E_f}), it propagates deeply in the CGM creating a hot, low density, ionized bipolar outflow that extends in the polar directions and regulates the inflow onto the galaxy, hence its growth. 

We now discuss more quantitatively the impact of feedback both on the MBH and the galaxy host, turning our attention to the relevance of the spin-dependent wind anisotropy.

\subsection{The impact of AGN feedback on the MBH}\label{sec: impact on MBH}
In order to understand how the MBH mass and spin evolve and to what extend their growth is influenced by AGN feedback, we need to understand its back-reaction on $f_\textrm{Edd}$, which characterizes the rate of such growth. 
\subsubsection{Evolution of $f_\textrm{Edd}$}\label{sec: fedd}
\begin{figure*}
    \centering
    \includegraphics[scale=0.5]{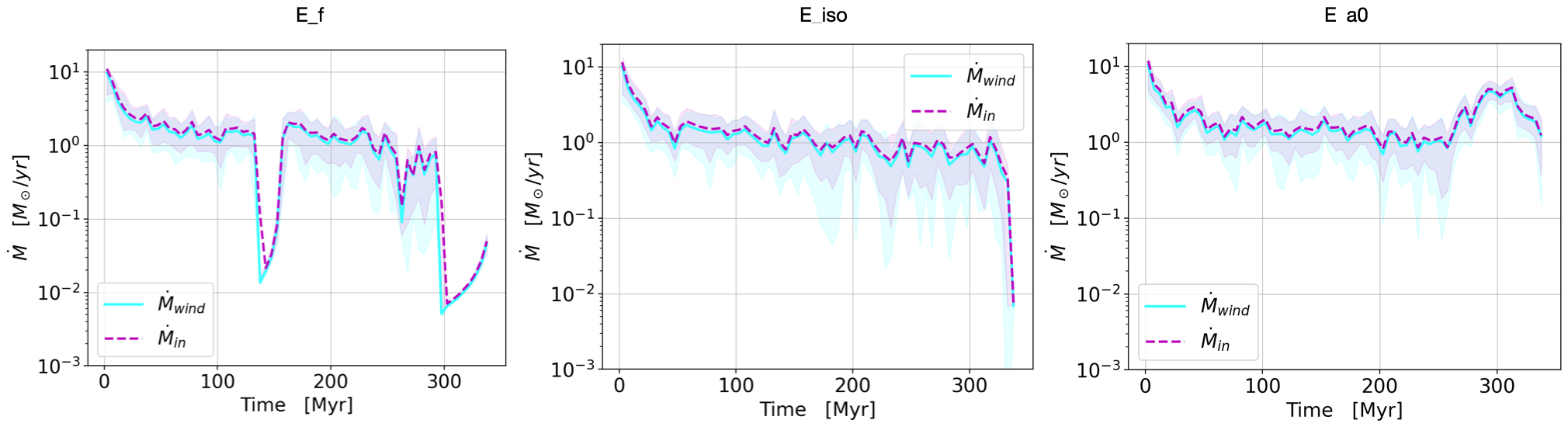}
    \caption{ The evolution of the mass inflow rate $\dot{M}_\textrm{in}$ on the sub-grid disc and the mass outflow rate $\dot{M}_\textrm{w}$ ejected in winds for the three {\labsim{E}} simulations. The solid and dashed lines correspond to the median values of these quantities over time bins of 5 Myr, while the shaded regions span from the 16th to the 84th percentiles of $\dot{M}_\textrm{in}$ and  $\dot{M}_\textrm{w}$ within the bins.}
    \label{fig: In&w}
\end{figure*}

In {\labsim{C}} simulations, the growth of MBH mass and spin is completely determined by the constrain $f_\textrm{Edd} = 1$, so it is independent on any effect feedback might have on the gas surrounding the MBH. Conversely, in {\labsim{E}} simulations, $f_\textrm{Edd}$ varies with time according to the evolution of the sub-grid disc+MBH system (Eq. \ref{Eq: fedd}), which is influenced by the action of feedback on the nuclear environment. Indeed, in this case, AGN feedback diminishes the gas density in the vicinity of the MBH, lowering the mass inflow rate on it and hence the AGN power. 

In order to have a better understanding of how $f_\textrm{Edd}$ evolves, we compute the time derivative $df_\textrm{Edd}/dt$ (see Appendix \ref{App: dfedd}), finding that it can be approximated as   
\begin{equation}
    \frac{d f_\textrm{Edd}}{dt} \simeq 5 f_\textrm{Edd}\Biggl(-C_\textrm{w}\frac{\dot{M}_\textrm{w}}{M_\alpha} + C_\textrm{in}\frac{\dot{M}_\textrm{in}}{M_\alpha} \Biggr),
    \label{Eq: dfedd}
\end{equation}
where $\dot{M}_\textrm{in}$ is the mass inflow rate on the sub-grid disc from resolved scales, as defined in \ref{App: inflow}, and the coefficients  $C_\textrm{w}$ and $C_\textrm{in}$ are both found to be $\simeq 0.1$, as detailed  in \ref{App: dfedd}.
Equation (\ref{Eq: dfedd}) describes the evolution of $f_\textrm{Edd}$ as driven by the disc mass loss, proportional to $\dot{M}_\textrm{w}$, and by the inflow $\dot{M}_\textrm{in}$ that supplies mass to the disc; in other words, the disc accretion rate is regulated by the balance between outflows and inflows. 
Because of the different signs in front of $C_\textrm{w}$ and $C_\textrm{in}$, the disc tends to evolve towards a stationary $df_\textrm{Edd}/dt \sim 0$ regime. Indeed, if the inflow exceeds the outflow, i.e. $C_\textrm{in}\dot{M}_\textrm{in} > C_\textrm{w}\dot{M}_\textrm{w}$, then the accretion rate $f_\textrm{Edd}$ increases and $\dot{M}_\textrm{w}$ grows as well (see Eq. \ref{Eq: mw-macc}). By contrast, if the disc mass consumption outpaces the external mass supply, $C_\textrm{w}\dot{M}_\textrm{w} > C_\textrm{in}\dot{M}_\textrm{in}$, the accretion rate decreases and hence $\dot{M}_\textrm{w}$ diminishes too. In this way inflows and outflows tend to adjust each other such that $C_\textrm{w}\dot{M}_\textrm{w} \sim C_\textrm{in}\dot{M}_\textrm{in}$, i.e. $df_\textrm{Edd}/dt \sim 0$.

\begin{figure}
    \centering
    \includegraphics[scale=0.5]{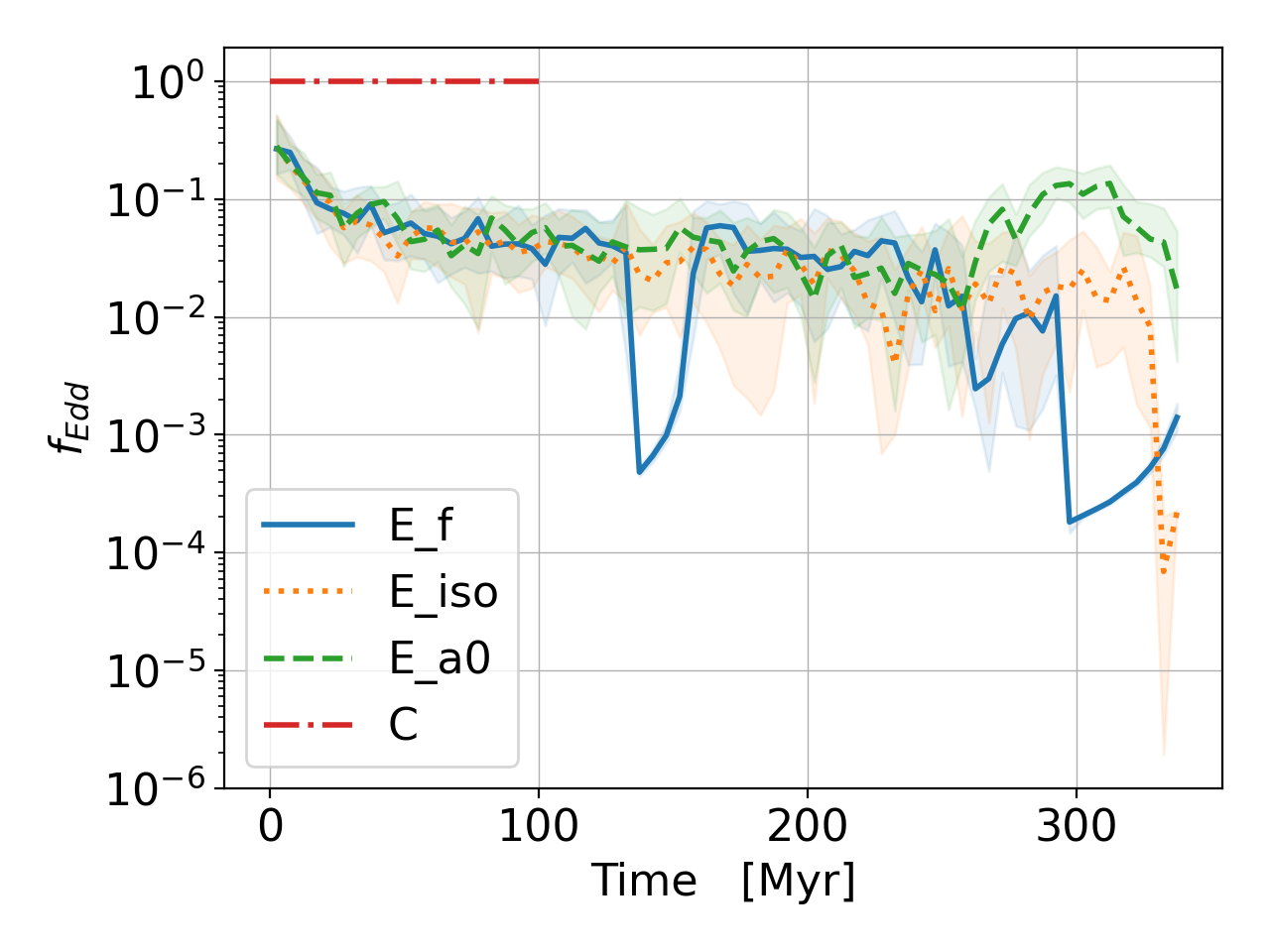}
    \caption{Evolution of $f_\textrm{Edd}$ in {\labsim{E}} and {\labsim{C}} simulations.The lines correspond to the median values of $f_\textrm{Edd}$ over time bins of 5 Myr, while the shaded regions represent the fluctuations within the bins from the 16-th to 84-th percentiles. } 

    \label{fig: fedd}
\end{figure}

Figure \ref{fig: In&w} shows the evolution of $\dot{M}_\textrm{w}$ and $\dot{M}_\textrm{in}$ in the three simulations {\labsim{E}} and reveals that in all cases $\dot{M}_\textrm{w} \sim \dot{M}_\textrm{in}$ and both quantities settle around $\sim 1 \rm\, M_\odot/yr$ after a transient of $\sim 50$ Myr. As a result, in all {\labsim{E}} simulations, $f_\textrm{Edd}$ decreases from its initial value $f_\textrm{Edd,0} = 1$ and, after $\sim 50$ Myr, attains a value $\sim 0.05$ (see Fig. \ref{fig: fedd}), which is consistent with observed Seyfert galaxies \citep{Ho09} and corresponds to a moderate luminous AGN with $L \simeq 6\cdot 10^{44}$ erg/s. 

\subsubsection{MBH mass and spin growth}
\begin{figure}
    \centering
    \includegraphics[scale=0.5, trim=0 0 50 0]{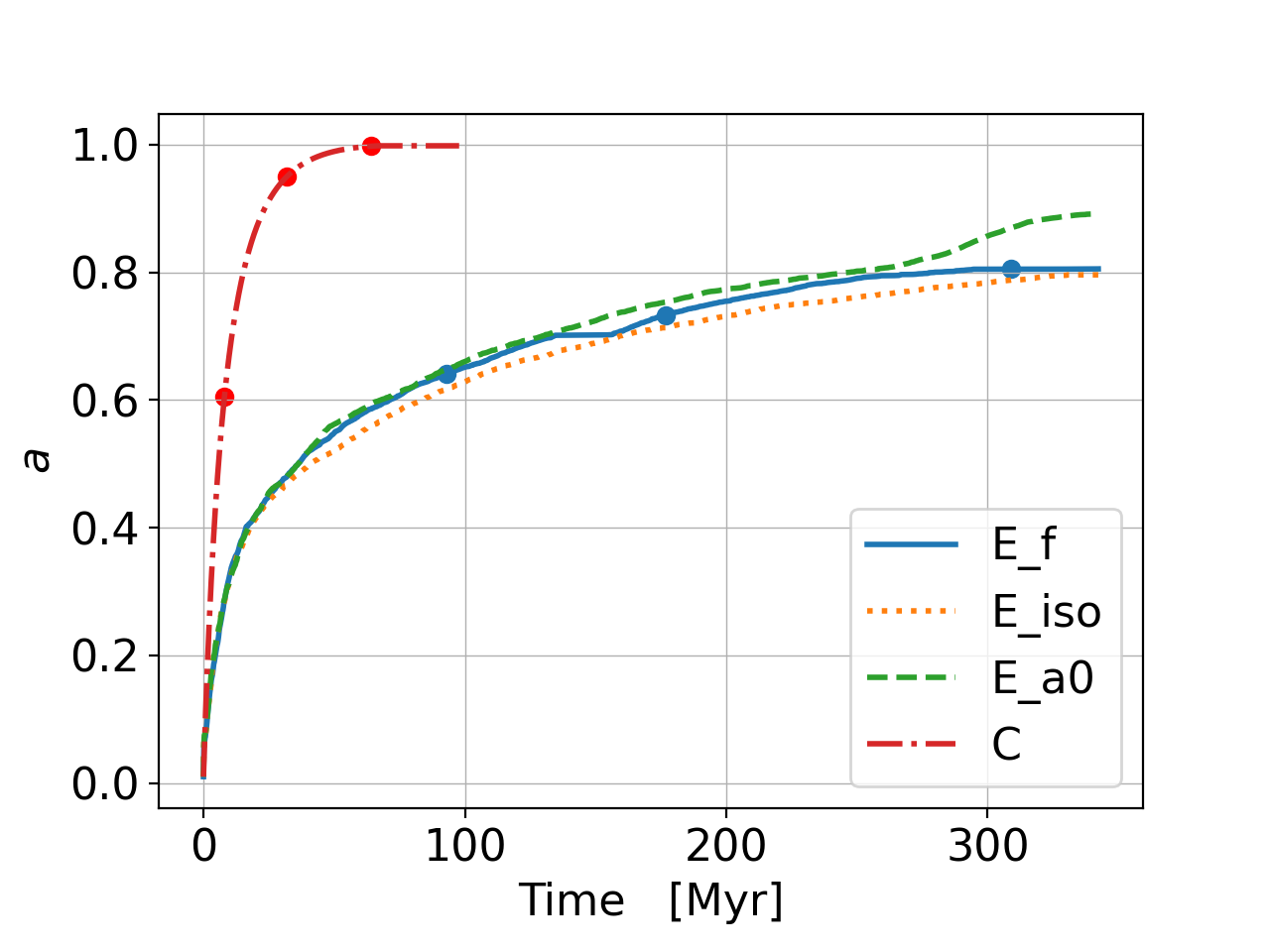}
    \includegraphics[scale=0.5]{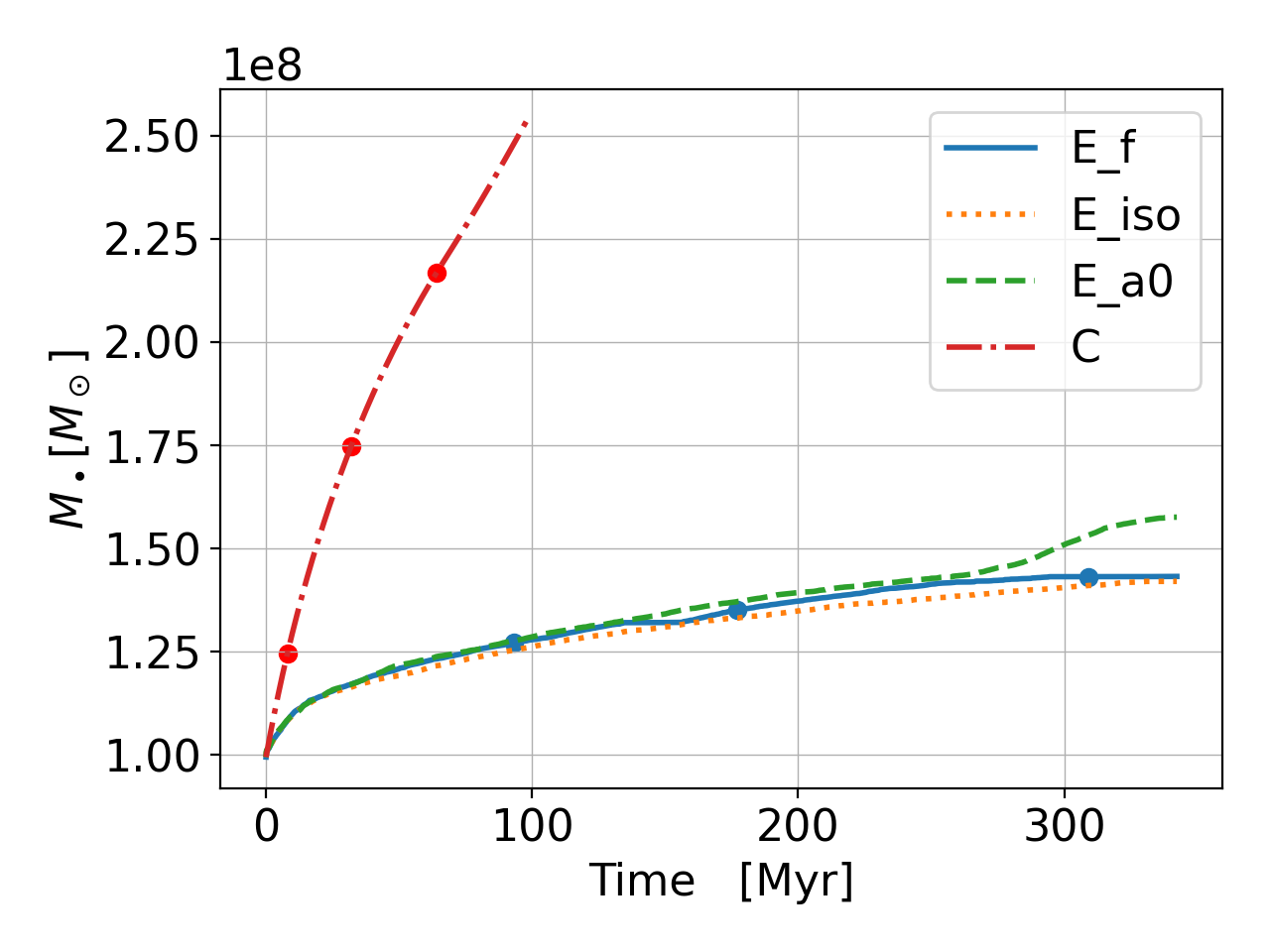}
    \caption{MBH spin and mass growth in {\labsim{E}} and {\labsim{C}} simulations. The red and blue bullets mark the instants corresponding to the snapshots shown in Figures~\ref{fig: C_f} and \ref{fig: E_f}.}
    \label{fig: Spin and Mass}
\end{figure}

The mass and spin evolution of the MBH in all our simulations is shown in Figure \ref{fig: Spin and Mass}. 
In {\labsim{C}} simulations, the growth of MBH mass and spin is constrained by $f_\textrm{Edd} = 1$, whereas in {\labsim{E}} simulations $f_\textrm{Edd}$ varies with time (Fig. \ref{fig: fedd}), according to the interplay between inflows and outflows, 
as discussed in \ref{sec: fedd}. In this second case $f_\textrm{Edd}$ lowers down to $\sim 0.05$, in this way the MBH mass and spin growth is severely delayed compared to the $f_\textrm{Edd} =1$ case. 

While the disc luminosity is directly linked to the MBH mass and spin growth rates (Eq. \ref{Eq:eta}), its angular pattern has a negligible impact on them. Nonetheless, in {\labsim{E}} simulations we notice a small dependence of these growth rates with the radiation angular pattern, i.e. the more isotropic the angular pattern is (where isotropy rises with {\labsim{a0}} $\to$ {\labsim{f}} $\to$ {\labsim{iso}}), the slower the mass and spin growth is. This can be attributed to the wind-ISM coupling increasing with isotropy (see section~\ref{sec: impact on host}), thus yielding smaller mass inflow rates on the MBH and hence lower AGN power.

\subsection{The impact of AGN feedback on the galaxy host}\label{sec: impact on host}

We now quantify the impact of AGN feedback on the host galaxy by probing the galaxy star formation history, the size of the central cavity and the AGN wind opening angle. 
\subsubsection{Star formation rate history}
\begin{figure*}
    \centering
    \includegraphics[scale=0.55]{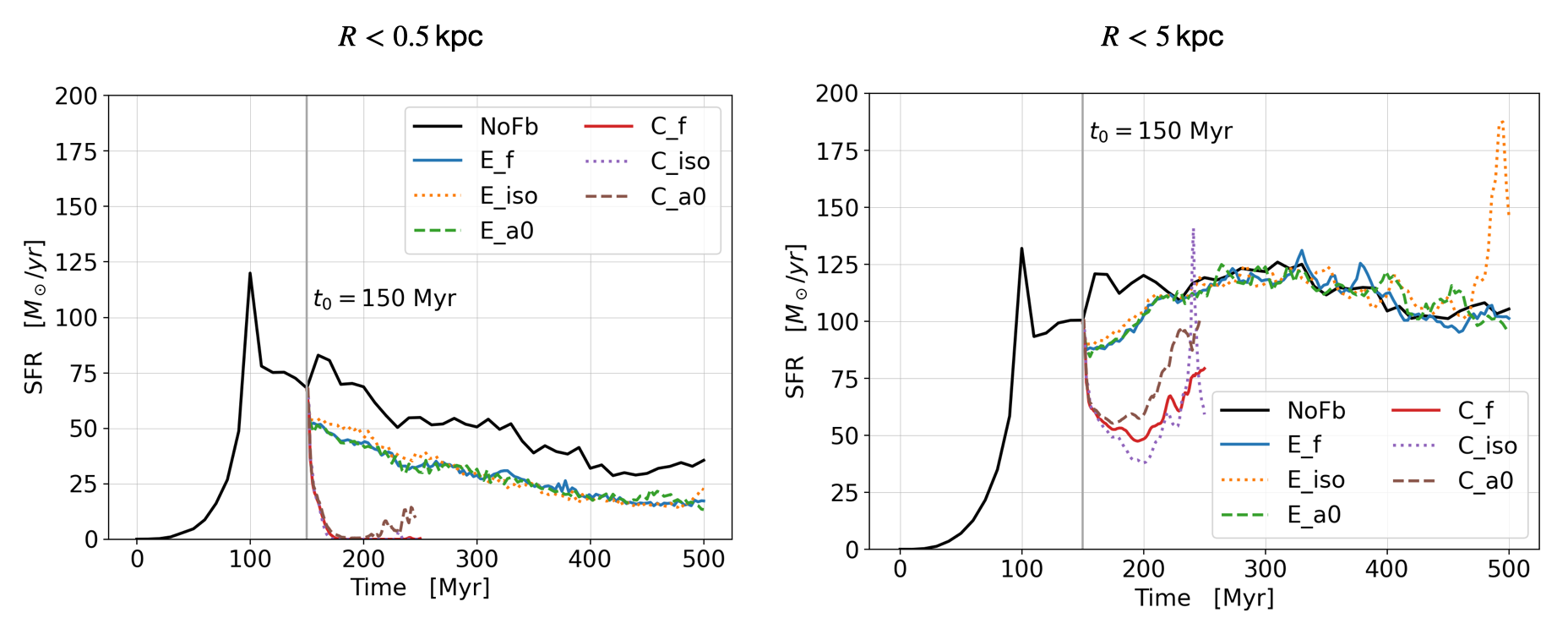}
    \caption{ Star Formation Rate (SFR) in the nuclear region (left) and at galactic scale (right) for {\labsim{E}}, {\labsim{C}} and {\labsim{NoFb}} simulations. 
    $t=0$ corresponds to the initial condition with the spinning gasous halo in hydrostatic equilibrium and $t_0 = 150$ Myr (when AGN feedback is turned on) is marked with a grey vertical line.    
    }
    \label{fig: SFH}
\end{figure*}
In Figure \ref{fig: SFH} we show the evolution of the star formation rate (SFR) with time. In order to differentiate the effect of the AGN feedback on nuclear and galactic scales, in the left panel we consider the SFR of the gas particles within 0.5 kpc from the MBH, while in the right panel the SFR of the gas particles within 5 kpc. In both panels we show results from the simulations sets {\labsim{C}} and {\labsim{E}} and from simulation {\labsim{NoFb}}, that with no feedback.

The {\labsim{NoFb}} simulation traces the evolution of the SFR starting from the initial condition with the spinning halo in hydrostatic equilibrium. At this time the SFR is zero, but as soon as the gaseous halo collapses, getting denser and cooler, the SFR grows and peaks at $\sim$ 100 Myr, both on nuclear and galactic scales. At this point, while in the central region the SFR starts to decline, in the whole galaxy stars continue to be formed at a rate of $100$ M$_\odot$/yr $\div 125$ M$_\odot$/yr. The switch-on of AGN feedback at $t_0 = 150$, in {\labsim{C}} and {\labsim{E}} runs, modifies these trends. 
In the central $0.5$ kpc, the SFR is reduced by a factor of $\lesssim 2$ in {\labsim{E}} simulations, while it is completely shut off in {\labsim{C}} simulations. The imprint of AGN feedback on galactic scales is less marked: in {\labsim{E}}-simulations the SFR within $5$ kpc is not appreciably altered, except for a $\lesssim 20$\% decrease during the first $\sim 50$ Myr of AGN activity, and a burst of SFR at the end of {\labsim{E\_iso}} due to a clump in a spiral arm, whereas it diminishes by a factor of $\lesssim 2$ in {\labsim{C}} runs. 
Overall, these trends do not show any significant dependence on the AGN radiation angular pattern, except for {\labsim{C}} simulations at galactic scale. In this case, we see that the more isotropic the luminosity angular pattern is, the more the SFR is suppressed, due to the increase of wind-ISM coupling. However, this appears as a second order effect and the impact of AGN feedback on the SFR is mainly driven by the AGN luminosity, rather than by its angular pattern.

\subsubsection{The size of the central cavity} \label{sec: cavity size}
\begin{figure}
    \centering
    \includegraphics[scale=0.5]{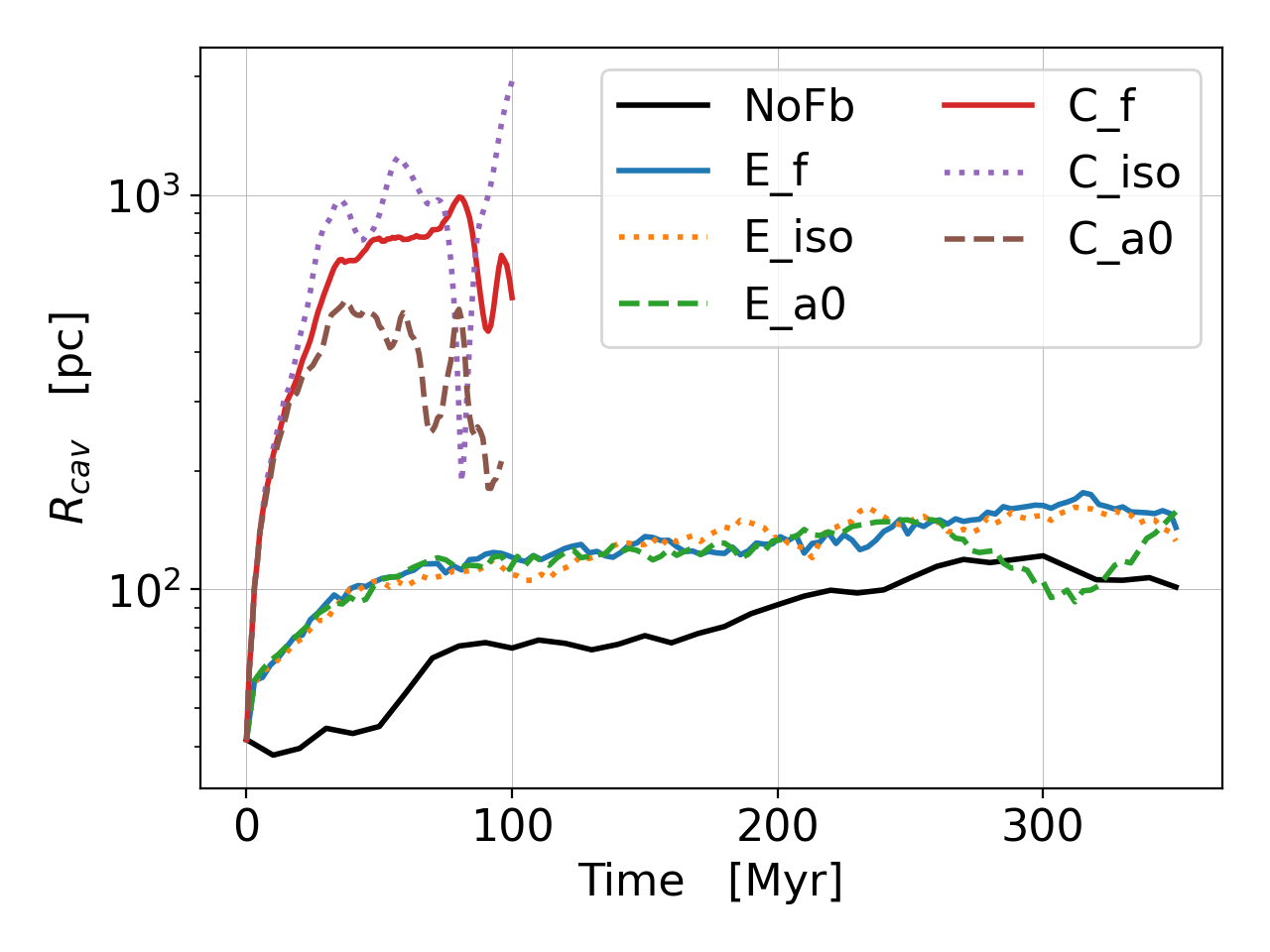}
    \caption{Evolution of the central cavity size in {\labsim{E}}, {\labsim{C}} and {\labsim{NoFb}} simulations.}
    \label{fig: Rcav}
\end{figure}
The trends seen in the SFR$(t)$ history are closely related to those seen in the size of the central cavity cleared by AGN feedback. Here the cavity size is measured as the radius $R_\textrm{cav}$ of the sphere centered in the MBH containing a gas mass equal to the MBH mass. Figure \ref{fig: Rcav} shows that in the simulation without feedback this quantity is initially (at $t_0$) about 50 pc and overall grows up to $\sim 100$ pc. In simulations with AGN feedback, AGN winds tends to push the gas away from the MBH, lowering the gas density in its surroundings and thus increasing $R_\textrm{cav}$. Such an increase is modest in {\labsim{E}} simulations, by a factor less than 2, while it is about an order of magnitude larger in {\labsim{C}} simulations. In this second case, we can distinguish a dependence of the cavity size with the feedback radiation angular pattern, where a more isotropic pattern results in a larger cavity. 

As for the SFR$(t)$, the cavity size seems to be determined at first order by the disc luminosity and only for high (Eddington) luminosities ({\labsim{C}} runs) by the radiation angular pattern, at least partially. In addition, the increase in the cavity size with disc luminosity and with disc radiation angular pattern mirrors what we found for the SFR, i.e., larger cavities are associated with lower SFRs. This suggests that star formation suppression in the galaxy nuclear region occurs via gas removal caused by AGN winds.

\subsubsection{The outflow opening angle}
\begin{figure}
    \centering
    \includegraphics[scale=0.5]{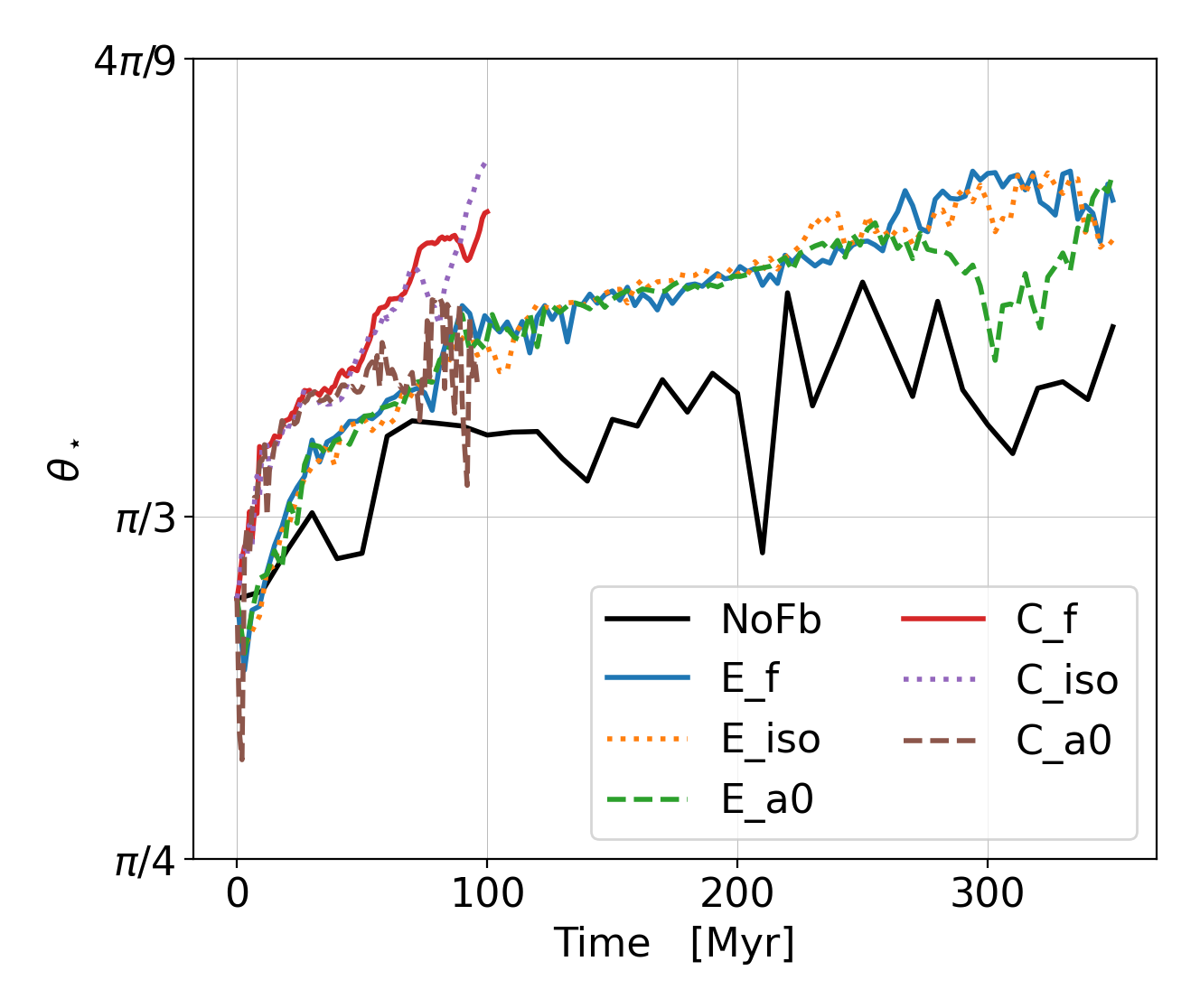}
    \caption{Evolution of the outflow semi-opening angle in {\labsim{E}}, {\labsim{C}} and {\labsim{NoFb}} simulations.}
    \label{fig: thetaSF}
\end{figure}
Finally, we measure the opening angle of the region within which the SFR is maintained below $1 \rm\,$ M$_\odot$/yr, which provides information about the anisotropy (or angular amplitude) of the bipolar outflow. 

In order to compute this quantity, we first determine, at any time $t$, the SFR angular profile $SFR(\theta; t)$, where $\theta$ is the polar angle from the galaxy axis. \footnote{We defined SFR$(\theta; t)$ as the sum of the SFR of all gas particles with polar coordinate $\theta _i \in [\theta -\Delta \theta/2 , \theta +\Delta \theta/2]$, or $\pi - \theta_i$ in the same interval, and within a distance of 5 kpc from the MBH. We used $\Delta \theta = \pi/40$.}
For all simulations,  SFR$(\theta; t) \simeq 10^{-2}\div 10^{-1}$ M$_\odot/$yr for small $\theta$, i.e. perpendicular to the galaxy disc plane, and rapidly increases at $\theta \gtrsim \pi/3$, reaching $\simeq 10 \div 10^2$ M$_\odot$/yr for $\theta \lesssim \pi/2$, i.e. in the galactic disc mid-plane. Given this trend, we define $\theta_\star$ as the angle below which the SFR$(\theta; t)< 1$ M$_\odot$/yr. Then, the larger $\theta_\star$, the wider the angular region where the outflow manages to keep the SFR below our treshold of 1 M$_\odot$/yr. In this sense, $\theta_\star$ measures the outflow semi-opening angle. 

In Figure \ref{fig: thetaSF}, we show how $\theta_\star$ evolves in our simulations and we notice a trend similar to that of SFR$(t)$ and $R_\textrm{cav}(t)$. Indeed, $\theta_\star(t) \gtrsim \pi/3$ in {\labsim{NoFb}} simulation, then becomes larger in {\labsim{E}} simulations and increases further in {\labsim{C}} runs. No particular trend of $\theta_\star$ is seen with the feedback radiation angular pattern, except for the final part of {\labsim{C}} simulations, which suggest that $\theta_\star$ is larger for more isotropic radiation angular patterns. Therefore, similarly to what we discussed for SFR$(t)$ and $R_\textrm{cav}(t)$, the disc luminosity, more than its angular pattern, determines the angular amplitude of the outflow, measured as the ability of AGN winds to hamper the SFR at large polar angles.

\section{Summary and Discussion}\label{sec: Discussion}

In this paper, we investigated the role that spin-dependent anisotropy of AGN winds \citep{Ishibashi19, Ishibashi20} has in shaping the evolution of MBHs and their host galaxies. To this purpose, we implemented in the code \textsc{gizmo} a sub-grid model for AGN feedback that takes into account the spin dependence of feedback anisotropy, linking and integrating existing modules for MBH accretion and spin evolution \citep{Cenci21} and AGN wind \citep{Torrey20}.
In doing so, we assumed that the AGN disc radiation couples with gas at the disc (sub-grid) scale, completely transferring its momentum. In this way, the nuclear wind that is launched inherits the luminosity angular pattern of the impinging radiation, which is set by the MBH spin. 
We initially tested our novel implementation by following the propagation of an AGN wind driven outflow into a homogeneous medium, and we compared the results against simple analytical models. 
Then, we considered an isolated galaxy setup, thought to be formed from the collapse of a spinning gaseous halo, and there we studied the impact of AGN feedback on the MBH and galaxy evolution. We considered different prescriptions for the MBH accretion rate, i.e., constant and equal to the Eddington rate or self-consistently evolved according to the resolved gas inflow onto the MBH. We also considered different degrees of anisotropy of the angular pattern of the launched AGN wind.

The most relevant results of our work can be summarised
as follows:
\begin{itemize}
    \item  MBH feedback and fueling are tightly intertwined. On the one hand, the AGN wind affects the gas reservoir that feeds the MBH and, on the other hand, gas flowing onto the MBH supplies material and power the wind. The disc-wind system is a complex, self-regulating system \cite{Fiore23}.  
    We found that accounting for such self-regulated evolution or not makes a crucial difference. In our simulations with an evolving MBH accretion rate, the AGN luminosity, initially set equal to Eddington $L_\textrm{Edd} = 1.2\cdot 10^{46}$ erg/s, drops down by $1\div 2$ orders of magnitude to $\simeq 6\cdot 10^{44}$ erg/s, exactly because the AGN feedback limits the inflow that powers itself. Such reduced luminosity corresponds to an Eddington factor $f_\textrm{Edd} \sim 0.05$, consistent with what estimated in Seyfert galaxies \citep{Ho09}. Such decrease in luminosity implies both a much slower MBH growth and a much weaker impact of the AGN on the host, when compared with simulations with constant Eddington MBH accretion rate. This highlights the importance of self-consistently evolving MBH accretion and feedback.

    \item 
    Once MBH accretion is allowed to evolve, we found that AGN feedback has a limited impact on the host galaxy, except for the central, $\simeq $kpc scale region. A smaller, $\simeq 100$ pc cavity is cleared around the AGN. At the same time the SFR is approximately halved within the inner $0.5$ kpc, while it remains at values typical of non active galaxies on larger scales. Put another way, our simulations indicate that isolated disc galaxies may be able to host luminous AGN activity without undergoing any significant star formation suppression on larger galaxy-scales. Our results agree with many observational studies that find no systematic signature of AGN feedback on the host galaxy SFR 
    \citep{Rosario13, Scholtz20, Smirnova22, Lammers22}.
    In addition, \cite{Lammers22} remarked that AGNs, despite not showing evidence for galaxy-wide quenching, have significantly suppressed central ($\sim$ kpc scale) SFR, lying up to a factor of 2 below those of the control non active  galaxies, in agreement with our findings.
    These results suggest that the integrated effect of secular AGN feedback, which is traced by the MBH mass, rather than an instantaneous AGN driven outflow, is required to significantly affect SF on galactic scales. In other words, the instantaneous AGN luminosity is not a proxy for the cumulative impact of AGN feedback on SF \citep{Bluck23}. In addition, we recall that besides AGN feedback, other physical processes can be responsible for star formation suppression in the nuclear region, e.g. the presence of bars \citep{Gavazzi15}.

\item The impact of AGN feedback on the host galaxy and on MBH growth is primarily determined by the AGN disc luminosity, rather than by its angular pattern. We found that MBHs accreting with constant Eddintong rate, corresponding to a bolometric luminosity of $L \sim 1.2 \cdot 10^{46}$ erg/s, are capable to clear kpc-scale cavities, suppressing SF by a factor of two on galactic scale, and driving outflows with large $\sim \pi/3 \div \pi/2$ semi-opening angles. For lower luminosities, $L\sim 6\cdot 10^{44}$, achieved once self-regulation is allowed, such effects are milder and restricted to the nuclear region. This is consistent with \cite{Torrey20} who found that increasing luminosity allows for further growth of the central cavity and suppression of SFR.

\item Conversely, the imprint of the AGN luminosity angular pattern on the MBH-galaxy evolution is less marked, and can be appreciated only in cases with high (Eddington) constant accretion rate, in which the AGN impact is overall stronger. 
For maximally anisotropic $\sim \cos\theta$ angular pattern (our {\labsim{a0}} simulations), most of the wind 
momentum and energy are funnelled in the MBH spin direction, i.e., perpendicularly to the galaxy disc, without much affecting the host galaxy. With more isotropic angular patterns, as occurring for higher MBH spin becauese of relativistic light bending, a larger fraction of the wind energy and momentum is distributed perpendicular to the spin, i.e., into the galactic disc, yielding a higher coupling between the wind and the galaxy ISM. Indeed, in our simulations with Eddington accretion, we observed that AGNs with isotropic luminosity more efficiently suppress the host SFR, up to a factor of two compared to the maximally anisotropic case, and more easily sweep away gas in the nuclear region, clearing cavities up to ten times larger. However, in simulations with smaller ($f_\textrm{Edd} \sim 0.05$) accretion rate, differences in the response of the galaxy to different AGN radiation anisotropies are negligible. As a consequence, we expect the spin-dependent anisotropy of AGN radiation to be relevant in those scenarios characterized by high and prolonged MBH accretion episodes and by high opening angle of the ISM disc as seen by the central MBH, as both features would increase the wind-galaxy coupling and make the galaxy response more sensitive to the radiation angular pattern. These conditions might be satisfied during galaxy mergers, where large amounts of gas are funnelled into the galactic nucleus, resulting in elevated MBH accretion rates and quasi-isotropic central gas geometries, or in high redshift galaxies, characterized by thick discs and by MBH accretion rates close to the Eddington limit over long periods of time \citep[e.g.][]{DiMatteo17,Barai19,Lupi19,Lupi22}.

\item The spin growth itself is influenced by the AGN angular pattern. Given a constant accretion rate, the spin growth naturally slows down due the distance of the ISCO from the MBH becoming smaller with increasing spin. In addition, as the spin becomes larger, both AGN luminosity and its isotropy increase and make the AGN feedback more capable to reduce the inflow on the MBH itself, further delaying its spin and mass growth. We witnessed a hint of this trend in our simulations with self regulated accretion, noting that more isotropic angular patterns yield slower MBH mass and spin growths. Although in our simulations this effect seems negligible.   
In this respect, we might speculate that, because of the angular pattern anisotropy, high-redshift slowly spinning MBHs might more easily attain accretion rates above the Eddington limit, 
as they would be less prone to alt accretion flows in the AGN disc equatorial plane via winds \citep{Lupi16}, but also because the efficiency of the jets potentially suppressing super-Eddington accretion rates is lower for lower MBH spins \citep{Regan19,Massonneau23}.

\end{itemize}

While in this paper we discussed the role of the spin-dependent anisotropy of AGN winds in the context of isolated disc galaxies, we remark that this effect might be crucial in other astrophysical scenarios where AGN feedback intervenes, such as the pairing and migration of MBH binaries \citep[e.g.,][]{DelValle18, Bollati23}.

Due to our simplified modeling, a number of caveats that we have
to keep in mind when interpreting our results do exist. Specifically:

\begin{itemize}
    \item Our accretion model does not include the geometrically thick, radiatively inefficient accretion mode that occurs below $ \lesssim f_\textrm{Edd} \sim 0.01$. For such low accretion rates the disc is still modelled as an $\alpha$-disc. Moreover, once the disc enters such low accretion regime, it becomes prone to launch a jet, a phenomenon not included in our model. Due to these limitations, we are not able to capture any transition from quasar to jet mode with its possible repercussion on the host galaxy and MBH evolution. Nonetheless, this should not occur frequently since, in our simulations, most of the time $f_\textrm{Edd}>0.01$.
    
    \item The coupling coefficient $\tau = \dot{M}_\textrm{w}v_\textrm{w}/(L/c)$ between AGN radiation and gas is assumed to be constant and equal to one. All the momentum transfer is assumed to take place at the disc (unresolved) scale with the launch of AGN wind. In a more realistic model 
    the radiation-gas coupling should evolve according to the ionization level of the gas, a fraction of the AGN radiation should be allowed to escape the unresolved disc scale and interact directly with the resolved ISM gas, exerting radiation pressure on it \citep{Costa18, Barnes18}. This would require performing Radiation-Hydrodynamics simulations, something we leave for future work.

    \item We did not model stellar winds and supernovae which, together with AGN feedback, contribute in driving galactic outflows, especially in dwarf galaxies \citep{Koudmani22}, and in regulating 
    the amount of gas present in the central region of a galaxy, thus further modulating the AGN fueling \citep{Dubois15, Alcazar17b}. These effects would add a further layer of complexity, beyond the scope of this paper, but nonetheless important in order to understand the detailed interaction between star formation, MBH growth and feedback and the overall co-evolution of MBHs and galaxies. 
    
    \item The multiphase structure of the ISM is not resolved, but evolved in a sub-grid fashion according to the \citep{Springel03} model. An explicit modelling of the
    inhomogeneous and clumpy ISM structure \citep{Hopkins18,Lupi19,Marinacci19} is beyond the scope of our novel investigation of spin-dependent feedback. 

\end{itemize}

\section*{Acknowledgements}

We acknowledge the CINECA award under the ISCRA initiative,
for the availability of high-performance computing resources and
support (project number HP10CAKAY4). The analyses reported in
this work have been mainly performed using pynbody \citep{pynbody}.

\section*{Data Availability}
The data underlying this article will be shared on reasonable request
to the corresponding author.


\bibliographystyle{aa}
\bibliography{example} 



\appendix

\section{BH accretion implementation} \label{App: acc implementation}
Here we review more in detail the sub-grid model for accretion and spin evolution we employed. In \cite{Cenci21}, the MBH particle is meant to represent a structured, sub-resolution system consisting of a MBH surrounded by an unresolved, warped accretion $\alpha$-disc \citep{SS73}. The MBH particle is completely characterized by its dynamical mass $M_{\bullet, \textrm{dyn}}$, \footnote{The dynamical mass is that used in the computation of the gravitational force.} whereas the sub-resolution, proper MBH by its mass $M_\bullet$ and the dimensionless spin-parameter $a$. The unresolved accretion disc is specified by its mass $M_\alpha$, its total angular momentum $\mathbf{J}_\alpha$, and its accretion rate $\dot{M}_\textrm{acc} = f_\textrm{Edd} \dot{M}_\textrm{Edd}$, where $\dot{M}_\textrm{Edd} = 4\pi G M_\bullet m_\textrm{p} / (\sigma_\textrm{T}\eta c)$ is the Eddington accretion rate, $m_\textrm{p}$ the proton mass,  $\sigma _\textrm{T}$ the Thomson scattering cross-section.

The time evolution of the MBH mass is governed by the accretion rate from the disc on the MBH $\dot{M}_\textrm{acc}$, whereas the mass of the unresolved disc evolves according to the mass inflow $\dot{M}_\textrm{in}$ from resolved scales, the mass outflow $\dot{M}_\textrm{w}$ and $\dot{M}_\textrm{acc}$:
\begin{align}
 & \dot{M}_\bullet = (1-\eta)\dot{M}_\textrm{acc}, \label{Eq: MBHdot} \\
 & \dot{M}_\alpha = \dot{M}_\textrm{in} - \dot{M}_\textrm{acc} - \dot{M}_\textrm{w}. \label{Eq: Malphadot}
\end{align}
Similarly, the MBH angular momentum $\mathbf{J}_\bullet = aGM_\bullet^2/c$ evolves due to the accretion from the disc and the Bardeen-Patterson torque $\mathbf{T}_\textrm{BP}$ \citep{BP75} that the MBH and the warped disc exert on each other. The disc angular momentum evolution $d\mathbf{J}_\alpha /dt$ is set equal and opposite to $d\mathbf{J}_\bullet /dt$, according to angular momentum conservation, plus terms that account for the exchange of angular momentum with the resolved environment through winds $\dot{\mathbf{J}}_\textrm{w}$ and inflows $\dot{\mathbf{J}}_\textrm{in}$:
\begin{align}
 & \frac{d\mathbf{J}_\bullet}{dt} = \textrm{sign}\bigl( \mathbf{J}_\bullet\cdot \mathbf{J}_\alpha\bigr)\Lambda _\textrm{ISCO}\dot{M}_\textrm{acc}  - \mathbf{T}_\textrm{BP} \label{Eq: dotJbh} \\
 & \frac{d\mathbf{J}_\alpha}{dt} = -\frac{d\mathbf{J}_\bullet}{dt} + \dot{\mathbf{J}}_\textrm{in} - \dot{\mathbf{J}}_\textrm{w}. \label{Eq: dotJalpha}
\end{align}
Here $\Lambda _\textrm{ISCO}$ is the specific angular momentum of the gas at the disc innermost stable circular orbit (ISCO) and $\mathbf{T}_\textrm{BP}$ is modelled as in \cite{Fiacconi18}. The angular momentum carried by the wind is given by 
\begin{equation}
    \dot{{\mathbf{J}}}_\textrm{w}dt = \mathbf{J}_\alpha \Biggl( 1 - \Bigl(1 -    \frac{\dot{M}_\textrm{w}dt}{M_\alpha }\Bigr)^{7/5} \Biggr),
    \label{Eq: Jw}
\end{equation}
 which is computed assuming that the wind arises as ejecta from the outskirt of the disc i.e.  $\dot{{\mathbf{J}}}_\textrm{w}dt = \int _{R'}^{R_\textrm{out}}\Sigma \sqrt{GM_\bullet R} 2\pi R\,dR$ where $R'$ satisfies $\dot{M}_\textrm{w}dt = \int _{R'}^{R_\textrm{out}} \Sigma 2\pi  R\,dR$, with $\Sigma \propto R^{-3/4}$ being the disc surface density and $R_\textrm{out}$ the disc external radius, defined as the radius where the disc becomes self-gravitating.

The accretion in the disc $\dot{M}_\textrm{acc}$ is self-consistently evolved according to the evolution of the sub-grid quantities determined by Eqs. (\ref{Eq: MBHdot})-(\ref{Eq: dotJalpha}). In particular, at any given MBH timestep $f_\textrm{Edd}$ is computed as 
\begin{equation}
    f_\textrm{Edd} \simeq 0.76 \Biggl( \frac{\eta}{0.1}\Biggr) \Biggl( \frac{M_\alpha}{10^4 M_\odot}\Biggr)^5 \Biggl(\frac{M_\bullet}{10^6 M_\odot} \Biggr)^{-47/7} \Biggl( \frac{a|\mathbf{J_\alpha}|}{3 |\mathbf{J}_\bullet|}\Biggr)^{-25/7},
    \label{Eq: fedd}
\end{equation}
see \citep{Fiacconi18} for its derivation. Below we discuss more in detail how the inflows $\dot{M}_\textrm{in}$ and $\dot{\mathbf{J}}_\textrm{in}$ are computed.

\subsection{Accretion from resolved scales on the subgrid disc} \label{App: inflow}

\begin{figure}
    \centering
    \includegraphics[scale=0.5]{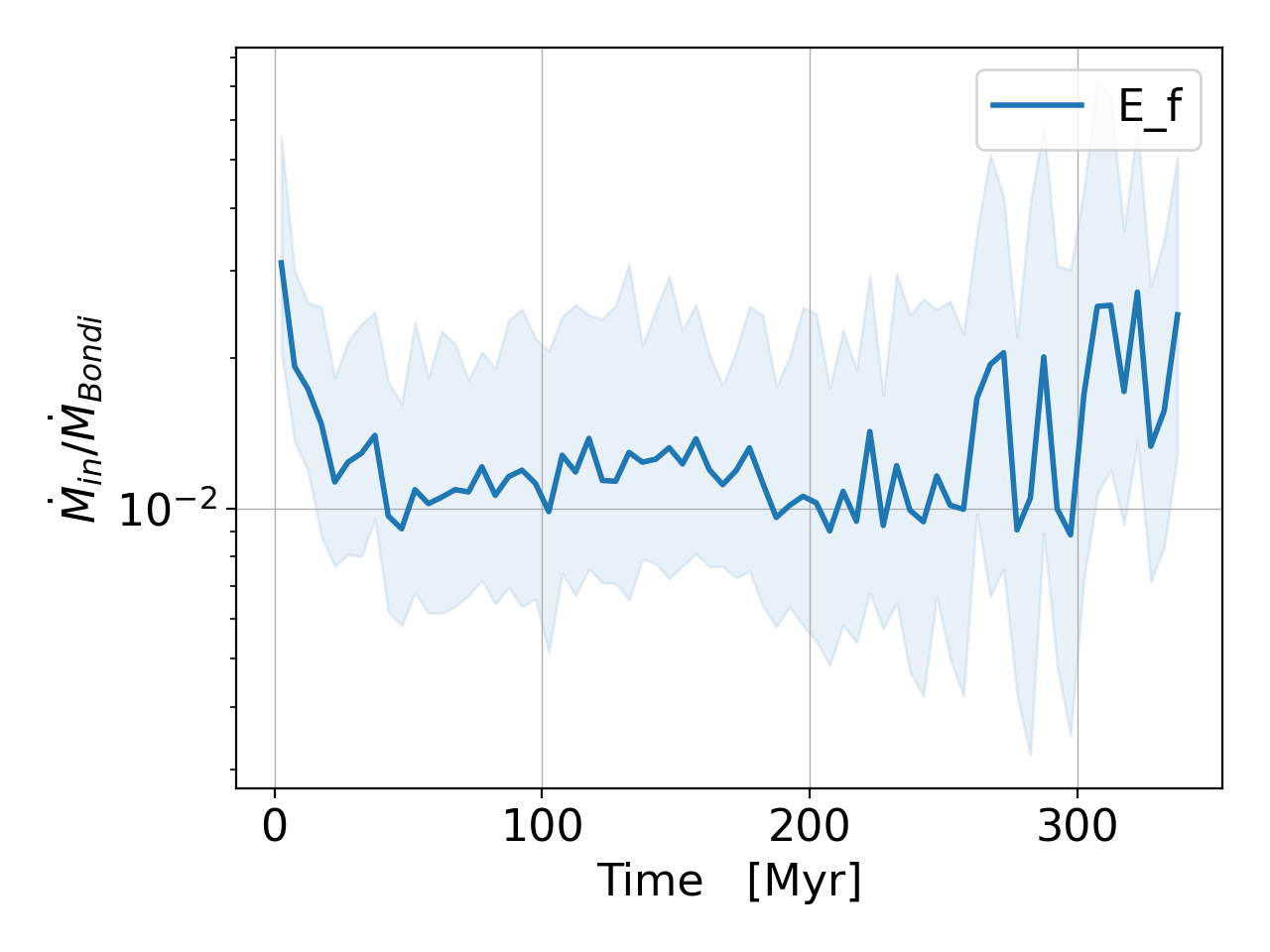}
    \caption{Ratio between the modidifed-Bondi accretion rate $\dot{M}_\textrm{in}$ and the classical Bondi rate in simulation {\labsim{E\_f}}.}
    \label{fig: acc_ratio}
\end{figure}

The accretion rate $\dot{M}_\textrm{in}$ from resolved scales on the sub-grid disc is estimated based on the properties of the gas particles within the MBH smoothing kernel $h_\bullet$, which is defined as a spherical region centered on the MBH enclosing a given effective number of particles $N_\textrm{ngb}$ capped to a maximum size $R_\textrm{max}$. 
In particular, the inflow $\dot{M}_\textrm{in}$ is estimated 
through a modified Bondi-Hoyle prescription that accounts for the angular momentum of gas in the BH kernel, as proposed by \cite{Tremmel16}. The argument suggested by \cite{Tremmel16}, similarly to the classical Bondi-Holye derivation, is based on the definition of a characteristic accretion radius $R_\textrm{acc}$, relative to the MBH, within which gas is bound to the MBH, and from which the accretion rate is computed as $\dot{M}_\textrm{in} \sim \pi R_\textrm{acc}^2 \rho v$,  where $v$ is the characteristic velocity of gas nearby the MBH and $\rho$ its the density. The accretion radius is defined as the radial distance
at which the gravitational potential of the MBH balances the internal and bulk energetics of the gas, but differently from the classical Bondi-Hoyle derivation, the computation is carried out in the reference frame of rotating gas, where the gas angular momentum
provides an effectively lower gravitational potential: 
$U_\textrm{eff}(r) = -GM_\bullet /r + \Lambda(r)^2/2r^2$, where $\Lambda(r)$ is the angular momentum per unit mass of the gas at distance $r$ from the MBH.
If the dominant motion of the gas is rotational rather than a
bulk flow, the energy balance reduces to the requirement that the effective potential balances with the thermal energy of the gas, i.e. $U_\textrm{eff} \sim c_{s}^2/2$. This allows to get $R_\textrm{acc}$ and then, assuming that the characteristic velocity of gas can be approximated as $v\sim c_\textrm{s}$, one obtains
\begin{equation}
    \dot{M}_\textrm{in} = \frac{4\pi (GM)^2 \rho c_\textrm{s}}{(v_\varphi ^2 + c_\textrm{s}^2)^2},
    \label{Eq: modified Bondi}
\end{equation}
where $v_\varphi \equiv \Lambda(r)/r$ encapsulates the amount of angular momentum support the gas has on the smallest resolved scales. Numerically, we compute $v_\varphi$ by estimating the specific angular momentum $\boldsymbol{\Lambda}$ of gas close to the MBH (i.e. prone to accrete) as the ratio between the kernel weighted averages of angular momentum and mass of particles in the MBH kernel. 
Then, assuming that angular momentum on the larger scales is conserved once the gas reaches the smallest resolved scale, we set  $v_\varphi = |\boldsymbol{\Lambda}| / R_\textrm{in}$, where $R_\textrm{in} \equiv h_\bullet/3$ is a proxy of the smallest resolved scale in the MBH kernel\footnote{In case $h_\bullet$ reaches it minimum value $\sim 2.8 \epsilon _\bullet$, we have that $R_\textrm{in}\sim \epsilon _\bullet$}. Then, if $v_\varphi < v_\textrm{bulk}$, where $v_\textrm{bulk}$ is the mass averaged velocity of gas in the MBH kernel, we adopt the usual Bondi-Hoyle formula  \citep{Springel05,DiMatteo05, Cenci21}
\begin{equation}
    \dot{M}_\textrm{in} = \frac{4\pi (GM)^2 \rho}{(v_\textrm{bulk} ^2 + c_\textrm{s}^2)^{3/2}},
    \label{Eq: Bondi calssical}
\end{equation}
otherwise Eq. (\ref{Eq: modified Bondi}).
Both in Eq. (\ref{Eq: modified Bondi}) and (\ref{Eq: Bondi calssical}) $\rho$ and $c_\textrm{s}$ are computed as mass-weighted averages on the gas particles within the MBH kernel. 
As an example, in Fig. \ref{fig: acc_ratio} we show the evolution of the ratio between $\dot{M}_\textrm{in}$ and the corresponding classical Bondi-Hoyle (Eq. \ref{Eq: Bondi calssical}) prescription in the simulation {\labsim{E\_f}}. This ratio is always much smaller than one, i.e. Eq. (\ref{Eq: Bondi calssical}) is never employed to compute the accretion on the sub-grid disc as gas kinematics is dominated by rotation instead of bulk motion. The Bondi-Hoyle accretion is about two orders of magnitude larger than the modified Bondi (Eq. \ref{Eq: modified Bondi}) accretion, in agreement with other works (e.g. \citealt{Hopkins11,Curtis16,Tremmel16, Akerman23}), which showed that the classical Bondi-Hoyle accretion has the tendency to overestimate the accretion on the MBH. Once the accretion rate $\dot{M}_\textrm{in}$ is computed, it is used to update the mass of the sub-grid disc, which evolves smoothly over time. Instead, the growth of the MBH particle, i.e. its dynamical mass $M_{\bullet, \textrm{dyn}}$, is performed in a discrete fashion by selecting stochastically the gas particles in the MBH kernel to be accreted, with probability $p \propto \max[M_\bullet + M_\alpha - (M_{\bullet, \textrm{dyn}} + \sum_k^N m_k), 0 ]$, where $m_k$ is the mass of the $k$-th gas particle among the $N$ selected. We remark that both in the computation of the accretion rate $\dot{M}_\textrm{in}$ and in the stochastic accretion, only non-wind particles are accounted, i.e. particles not spawned from the sub-grid disc (see section~\ref{Sec: agn wind}), since such particles are initialized as outflowing and hence they do not contribute to the accretion flow on the MBH. 

From the accretion $\dot{M}_\textrm{in}$ on the sub-grid disc, we easily get the accreted angular momentum on the sub-grid disc as  $\dot{\mathbf{J}}_\textrm{in} = \dot{M}_\textrm{in} \boldsymbol{\Lambda} dt$, where  $| \boldsymbol{\Lambda} |$ is limited as in \cite{Cenci21} in order to account for not resolved processes like shocks and torques that make the accreting gas further loose angular momentum before accreting on the sub-grid disc at not resolved scales.  

While the sub-grid disc mass is capped in order to avoid it from becoming self-gravitating, it is allowed to become zero, i.e. quiescent MBH, due to the accretion on the MBH and wind ejection. Once $M_\alpha$ vanishes, the disc is refilled with probability $p = (\sum \dot{M}_\textrm{in} dt)/M_\textrm{seed}$, where $\sum \dot{M}_\textrm{in} dt$ is the mass inflow collected since the MBH became quiescent and $M_\textrm{seed}$ is a free parameter ($10^5 M_\odot$ in our simulations). If refilling occurs, the new sub-grid disc is initialized with $M_\alpha = \max[M_\textrm{seed},\sum \dot{M}_\textrm{in} dt]$, $f_\textrm{Edd} = 0.1$, angular momentum set by Eq. (5) of \cite{Cenci21} and its direction equal to that of the angular momentum in the MBH kernel.

\subsection{The derivative of $f_\textrm{Edd}$} \label{App: dfedd}
\begin{figure}
    \centering
    \includegraphics[scale=0.5]{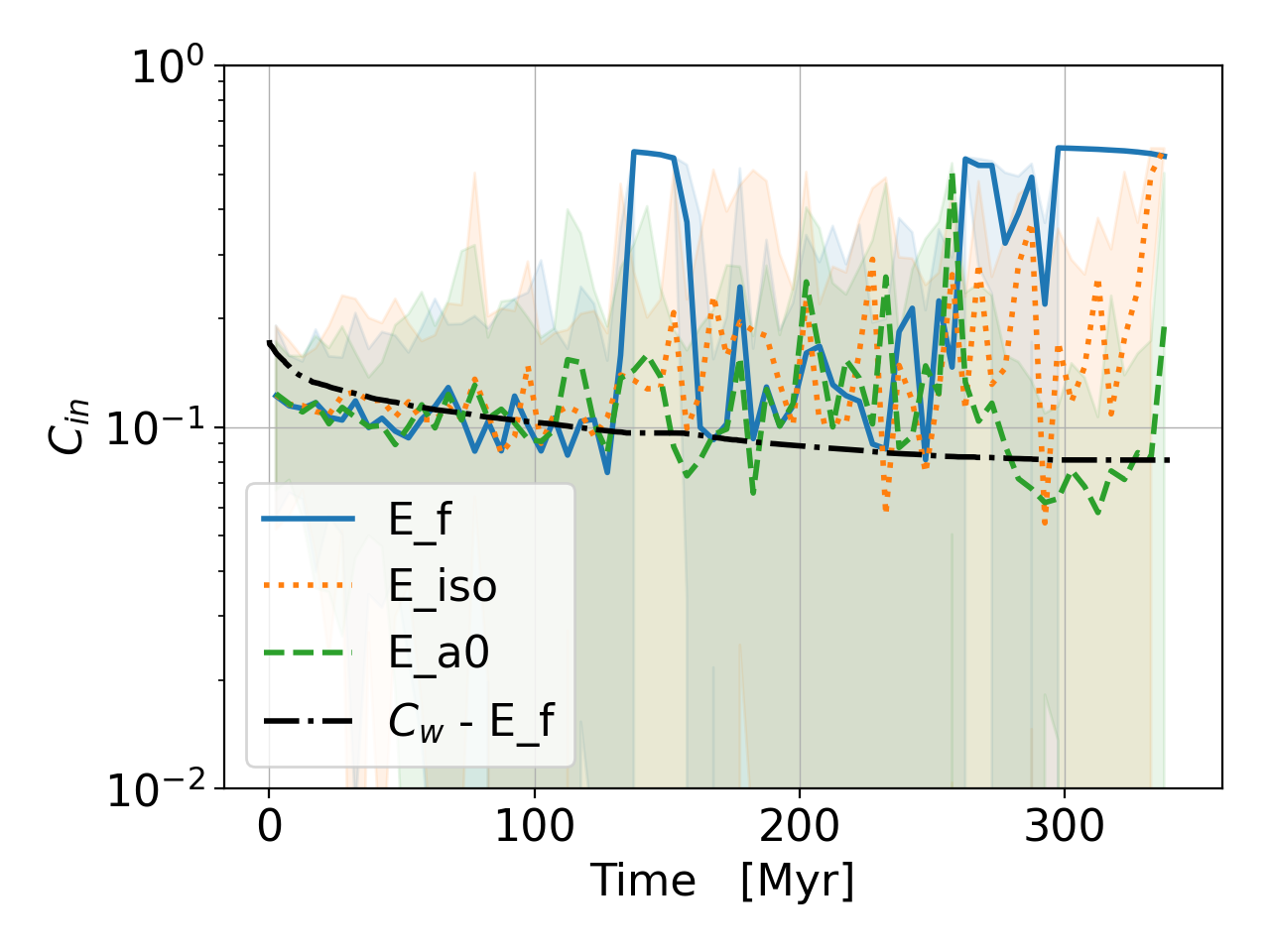}
    \caption{Coefficients $C_\textrm{in}$ in {\labsim{E}} simulations. The lines correspond to the median values over time bins of 5 Myr, while the shaded regions span from the 16th to the 84th percentiles over these bins.  The coefficient $C_\textrm{w}$ is also shown for the {\labsim{E\_f}}  run. 
    }
    \label{fig: Cin}
\end{figure}

Here we compute the derivative of $f_\textrm{Edd}$ and find an approximate expression suitable for interpreting results shown in section~\ref{sec: impact on MBH}. Starting from Eq.~(\ref{Eq: fedd}) and replacing $a/J_\bullet = c/GM_\bullet^2$, the time derivative of $f_\textrm{Edd}$ reads
\begin{equation}
\begin{aligned}
    \frac{df_\textrm{Edd}}{dt} &= f_\textrm{Edd} \Biggl(\frac{\dot{\eta}}{\eta} + 5 \frac{\dot{M}_\alpha}{M_\alpha} + \frac{3}{7}\frac{\dot{M}_\bullet}{M_\bullet} - \frac{25}{7}\frac{\dot{J}_\alpha}{J_\alpha} \Biggr) \simeq \\
    & \simeq f_\textrm{Edd} \Biggl(\frac{\dot{\eta}}{\eta} + 5 \frac{\dot{M}_\textrm{in}-\dot{M}_\textrm{acc}-\dot{M}_\textrm{w}}{M_\alpha} + \frac{3}{7}\frac{(1-\eta)\dot{M}_\textrm{acc}}{M_\bullet} \dots\\
    & \hspace{0.5\linewidth} \dots  - \frac{25}{7}\frac{\dot{J}_\textrm{in} -\dot{J}_\bullet -\dot{J}_\textrm{w} }{J_\alpha} \Biggr),
    \label{Eq: dfedd1}
\end{aligned}    
\end{equation}
where in the second row, we plugged in Eqs.~(\ref{Eq: MBHdot}, \ref{Eq: Malphadot}), and \eqref{Eq: dotJalpha} in place of $\dot{M}_\alpha$, $\dot{M}_\bullet$ and $\dot{J}_\alpha$ and, for Eq.~\eqref{Eq: dotJalpha}, we approximated all vectors as lying along the same direction. Eq.~\eqref{Eq: dfedd1} can be simplified as follows. 
First, we note that $-5\dot{M}_\textrm{w}/M_\alpha + 25/7 \dot{J}_\textrm{w}/J_\alpha = 0$, according to the definition of $\dot{J}_\textrm{w}$ (Eq. \ref{Eq: Jw}). In other words, the removal of mass and angular momentum from the disc due to the wind ejection does not affect the disc accretion rate. 
In addition, $3/7(1-\eta)\dot{M}_\textrm{acc}/M_\bullet$ and $\dot{\eta}/\eta \propto \dot{M}_\textrm{acc}/M_\bullet$ are both $\ll \dot{M}_\textrm{acc}/M_\alpha$, being $M_\alpha \ll M_\bullet$. Similarly, noting that $\mathbf{T}_\textrm{BP}$ in Eq. (\ref{Eq: dotJbh}) doesn't affect affect the spin modulus but only its direction, we can write $|\dot{J}_\bullet/J_\alpha| = \Lambda_\textrm{ISCO}/(J_\alpha/M_\alpha) \cdot \dot{M}_\textrm{acc}/M_\alpha \ll \dot{M}_\textrm{acc}/M_\alpha$ since the specific angular momentum of gas orbiting at the ISCO is much smaller than the disc total specific angular momentum $J_\alpha/M_\alpha$. In this way, Eq. (\ref{Eq: dfedd1}) reduces to
\begin{equation}
    \frac{df_\textrm{Edd}}{dt} \simeq f_\textrm{Edd} \Biggl( -5 \frac{\dot{M}_\textrm{acc}}{M_\alpha} +5 \frac{\dot{M}_\textrm{in}}{M_\alpha} - \frac{25}{7}\frac{\dot{J}_\textrm{in}}{J_\alpha} \Biggr). 
\end{equation}
Now, if we write $\dot{J}_\textrm{in} = \langle \Lambda \rangle \dot{M}_\textrm{in}$ (see Section \ref{App: inflow}) and  $\dot{M}_\textrm{acc} = \eta_\textrm{w}^{-1} \dot{M}_\textrm{w}$ (using Eq. \ref{Eq: mw-macc}), we obtain 
\begin{equation}
    \frac{d f_\textrm{Edd}}{dt} \simeq 5 f_\textrm{Edd}\Biggl(-\frac{1}{\eta_\textrm{w}}\frac{\dot{M}_\textrm{w}}{M_\alpha} + \Biggl(1 -\frac{5}{7}\frac{\langle \Lambda \rangle}{J_\alpha/M_\alpha} \Biggr)\frac{\dot{M}_\textrm{in}}{M_\alpha} \Biggr).
    \label{Eq: dfedd3}
\end{equation}
This equation points out that the leading terms driving the evolution of $f_\textrm{Edd}$ are the loss of mass due to accretion, here expressed in terms of $\dot{M}_\textrm{w}$, and the replenishment of mass from resolved scales, represented by $\dot{M}_\textrm{in}$. If we define the coefficients $C_\textrm{w} = \eta_\textrm{w}^{-1}$ and $C_\textrm{in} = (1 -5\langle \Lambda \rangle/7(J_\alpha/M_\alpha))$ we get Eq. (\ref{Eq: dfedd}). 
For our choice of parameters, being $v_\textrm{w}/c = 0.01$, we have that $C_\textrm{w} = 0.1 (\eta/0.1)^{-1}$ is of the same order of $C_\textrm{in}$, whose value is shown in Fig. \ref{fig: Cin} for the three simulations {\labsim{E}}. 

\label{lastpage}
\end{document}